\documentclass[apl,twocolumn,showpacs,groupedaddress, longbibliography,10pt]{revtex4-1}

\usepackage{amsmath,amssymb}
\usepackage{graphicx}
\usepackage{pgfplots}
\usepackage{subfigure}
\usepackage{xcolor}
\usepackage{tikz}
\RequirePackage{atbegshi}
\usetikzlibrary{positioning}
\usetikzlibrary{arrows}
\usetikzlibrary{decorations.text}

\usetikzlibrary{decorations.pathmorphing}

\setcitestyle{super}

\begin{document}
\title{Interaction between light and highly confined hypersound \\ in a silicon photonic nanowire}

\author{Rapha\"{e}l Van Laer}
\email{raphael.vanlaer@intec.ugent.be}
\author{Bart Kuyken}
\author{Dries Van Thourhout and Roel Baets}
\affiliation{Photonics Research Group, Ghent University--imec, Belgium \\ Center for Nano- and Biophotonics, Ghent University, Belgium \vspace{-4mm}}

\maketitle

\textbf{In the past decade, there has been a surge in research at the boundary between photonics and phononics \cite{Maldovan2013}. Most efforts centered on coupling light to motion in a high-quality optical cavity \cite{Kippenberg2008a}, typically geared towards observing the quantum state of a mechanical oscillator \cite{Chan2011b}. It was recently predicted that the strength of the light-sound interaction would increase drastically in nanoscale silicon photonic wires \cite{Rakich2012}. Here we demonstrate, for the first time, such a giant overlap between near-infrared light and gigahertz sound co-localized in a small-core silicon wire. The wire is supported by a tiny pillar to block the path for external phonon leakage, trapping  $\mathbf{10} \; \textbf{GHz}$ phonons in an area below $\mathbf{0.1 \; \boldsymbol\mu}\textbf{m}^{\mathbf{2}}$. Since our geometry can be coiled up to form a ring cavity, it paves the way for complete fusion between the worlds of cavity optomechanics and Brillouin scattering. The result bodes well for the realization of low-footprint optically-pumped lasers/sasers\cite{Grudinin2010} and delay lines\cite{Zhu2007b} on a densely integrated silicon chip.}

The diffraction of light by sound was first studied by L\'{e}on Brillouin in the early 1920s. Therefore such inelastic scattering has long been called \textit{Brillouin scattering}\cite{Boyd2008}. On the quantum level, the process annihilates pump photons while creating acoustic phonons and red-shifted Stokes photons. The effect is known as \textit{stimulated} Brillouin scattering (SBS) when the sound is generated by a strong modulated light field. This sets the stage for a self-sustaining feedback loop: the beat note between two optical waves (called the \textit{pump} and the \textit{Stokes}) generates sound that reinforces the initial beat note.

In a seminal experimental study\cite{Chiao1964}, Brillouin scattering was viewed as a source of intense coherent sound. Later, the effect became better known as a noise source in quantum optics \cite{Shelby1985} and for applications such as spectrally pure lasing \cite{Grudinin2009,Tomes2009b,Lee2012a}, microwave signal processing \cite{Pant,Li2013a}, slow light\cite{Thevenaz2008}, information storage \cite{Zhu2007b} and phononic band structure mapping \cite{Gorishnyy2005}.


Traditionally \cite{Kobyakov2009,Shelby1985,Grudinin2010,Boyd2008,Chiao1964,Bahl2011a,Pant,Li2013a,Gorishnyy2005,Grudinin2009,Tomes2009b,Bahl2012,Lee2012a,Zhu2007b,Thevenaz2008,Eggleton2013},    the photon-phonon interaction was mediated by the material nonlinearity. Electrostriction drove the phonon creation, and phonon-induced permittivity changes lead to photon scattering. This conventional image of SBS as a bulk effect, without reference to geometry, breaks down in nanoscale waveguides. The impressive progress in engineering radiation pressure in micro- and nanoscale systems\cite{Li2008,Li2009d,Roels2009a,Wiederhecker2009,VanThourhout2010a} recently inspired the theoretical prediction of enormously enhanced photon-phonon coupling\cite{Rakich2012,Qiu2013,Okawachi2012,VanLaer2014} in silicon nanowires. In such waveguides, boundary effects can no longer be neglected. Thus both electrostriction and radiation pressure create phonons. Equivalently, the new theory takes into account not only bulk permittivity changes but also the shifting material boundaries. The strong photon confinement offered by these waveguides boosts both types of optical forces. However, destructive interference between the two contributions may still completely cancel the photon-phonon coupling. The giant light-sound overlap arises exclusively when both bulk and boundary forces align with the phonon field \cite{Rakich2012,Qiu2013}.

Unfortunately, typical silicon-on-insulator wires provide only weak phonon confinement because there is little elastic mismatch between the silicon core and the silicon dioxide substrate. The large coupling strength was thus thought to be only accessible in silicon wires that are fully suspended in air\cite{Rakich2012,Qiu2013,Okawachi2012,VanLaer2014}. This requirement severely compromises the ability to create centimeter-scale interaction lengths, which are paramount to reduce the required pump power for SBS.

In this work, we take the middle ground between these conflicting requirements. By partially releasing a silicon wire from its substrate, we drastically improve phonon confinement (fig.1a-c). There is still some phonon leakage through the pillar, but it is sufficiently low to tap the large overlap between the optical forces and the hypersonic mode (fig.1d). Moreover, it is straightforward to increase the interaction length in our design. Building on this compromise, we demonstrate an order-of-magnitude performance leap in the photon-phonon coupling.

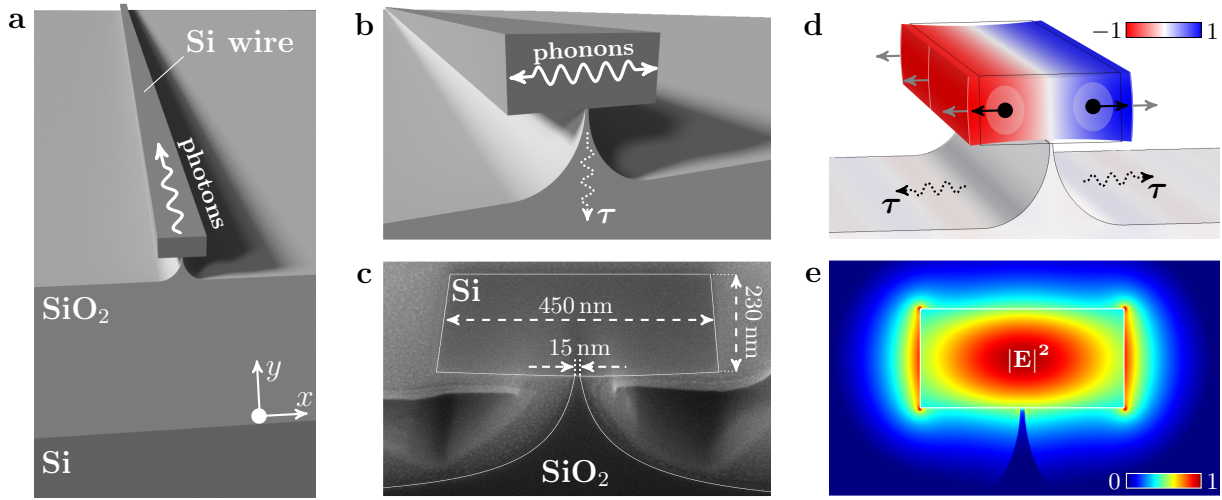
\begin{figure*}
\centering
\label{fig:1}
\subfigure{   
		\begin{tikzpicture}[>=stealth']
		
   		 \node[anchor=south west,inner sep=0] (image) at (0,0) {\raisebox{0pt}{\includegraphics[scale=0.257]{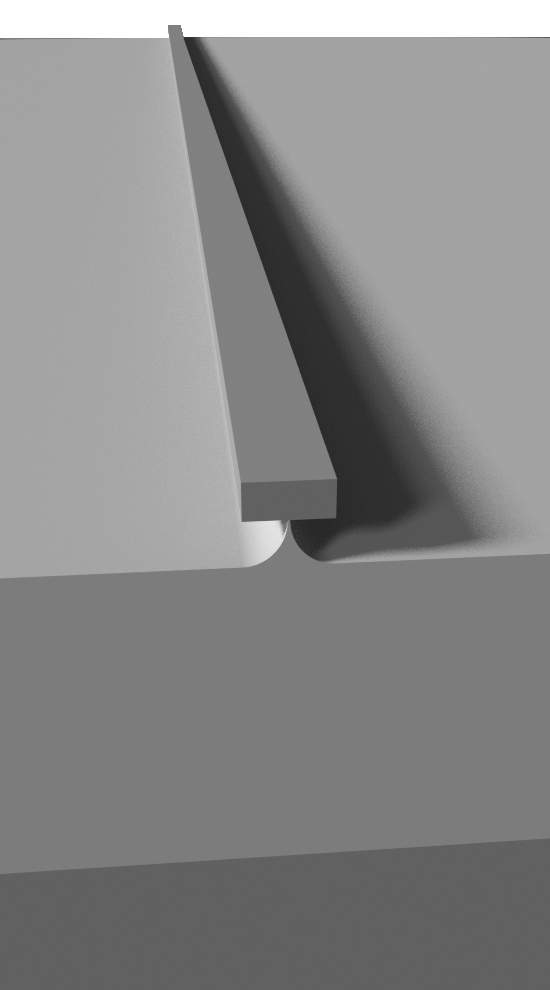}}};
    		\begin{scope}[x={(image.south east)},y={(image.north west)}]
       		\node at (0.08,0.07) {\textcolor{white}{\large{\textbf{Si}}}};
		\node at (0.73,0.9) {\textcolor{white}{\large{\textbf{Si wire}}}};
		\draw[white] (0.39,0.8) -- (0.53,0.87);
		\draw[white,->,very thick,decorate,decoration={snake,amplitude=.8mm,segment length=3mm,post length=2mm}] (0.52,0.52) -- (0.44,0.70);
		\node[rotate=-70] at (0.60,0.62) {\textcolor{white}{{\textbf{photons}}}};
		\node at (-0.06,0.94) {\large{\textbf{a}}};
		\node at (0.15,0.37) {\textcolor{white}{\large{\textbf{$\text{SiO}_{2}$}}}};
		\draw[white,*->,thick,rotate around={3:(0.77,0.16)}] (0.77,0.16) -- (0.97,0.16);
		\draw[white,->,thick,rotate around={93:(0.80,0.16)}] (0.80,0.16) -- (1.0,0.16);
		\node at (0.96,0.195) {\textcolor{white}{\large{{${x}$}}}};
		\node at (0.85,0.25) {\textcolor{white}{\large{{${y}$}}}};
    		\end{scope}
		\end{tikzpicture}
}
\subfigure{   
		\begin{tikzpicture}[>=stealth']
		
   		 \node[anchor=south west,inner sep=0] (image) at (0,0) {\raisebox{0pt}{\includegraphics[width=0.6\columnwidth]{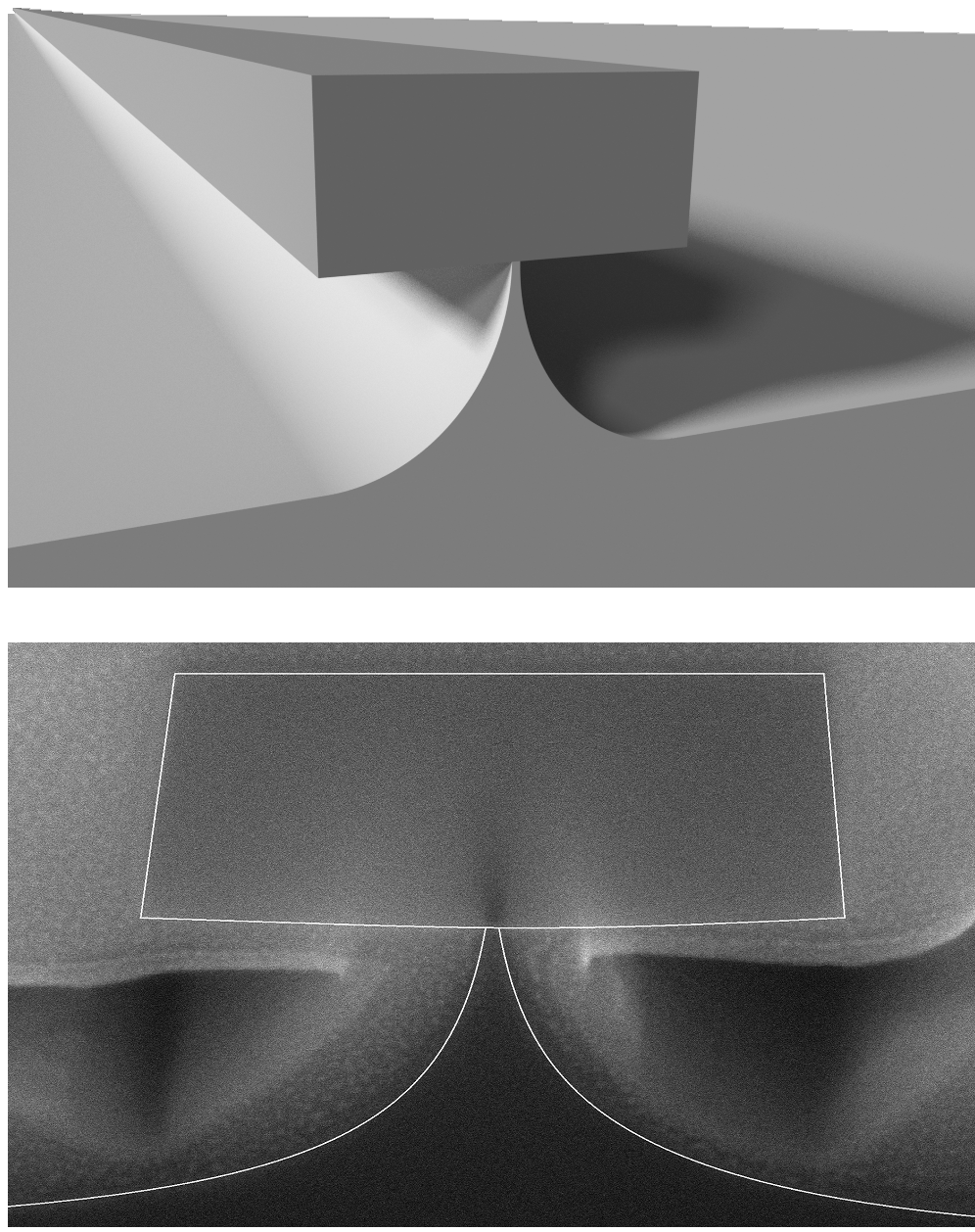}}};
    		\begin{scope}[x={(image.south east)},y={(image.north west)}]
		\draw[white,<->,very thick,decorate,decoration={snake,amplitude=1.2mm,segment length=3mm,pre length=3mm,post length=2mm}] (0.33,0.85) -- (0.71,0.87);

		\draw[white,->,thick,densely dotted,decorate,decoration={snake,amplitude=0.8mm,segment length=2.5mm,pre length=2mm,post length=2mm}] (0.53,0.74) -- (0.53,0.56);
		\node at (0.58,0.57) {\textcolor{white}{\large{{$\mathbf{\boldsymbol\tau}$}}}};

		\node[rotate=3] at (0.52,0.905) {\textcolor{white}{{\textbf{phonons}}}};

		\node at (0.5,0.05) {\textcolor{white}{\large{\textbf{$\text{SiO}_{2}$}}}};
		\node at (0.22,0.421) {\textcolor{white}{\large{\textbf{Si}}}};
		
		\draw[white,<->,dashed,thick] (0.170,0.36) -- (0.852,0.36);
		\node at (0.5,0.39) {\textcolor{white}{{$450 \, \text{nm}$}}};
		
		\draw[white,densely dotted,thick] (0.498,0.247) -- (0.498,0.282);
		\draw[white,densely dotted,thick] (0.511,0.247) -- (0.511,0.282);
		\draw[white,->,densely dashed,thick] (0.38,0.265) -- (0.495,0.265);
		\draw[white,<-,densely dashed,thick] (0.514,0.265) -- (0.629,0.265);
		\node at (0.51,0.305) {\textcolor{white}{{$15 \, \text{nm}$}}};
		
		\draw[white,<->,dashed,thick] (0.91,0.255) -- (0.91,0.452);
		\draw[white,densely dotted] (0.87,0.256) -- (0.905,0.256);
		\draw[white,densely dotted] (0.851,0.452) -- (0.905,0.452);
		\node[rotate=-90] at (0.955,0.355) {\textcolor{white}{{$230 \, \text{nm}$}}};
		
		\node at (-0.04,0.96) {\large{\textbf{b}}};
		\node at (-0.04,0.45) {\large{\textbf{c}}};
    		\end{scope}
		\end{tikzpicture}
}
\subfigure{   
		\begin{tikzpicture}[>=stealth']
		
   		 \node[anchor=south west,inner sep=0] (image) at (0,0) {\raisebox{0pt}{\includegraphics[width=0.6\columnwidth]{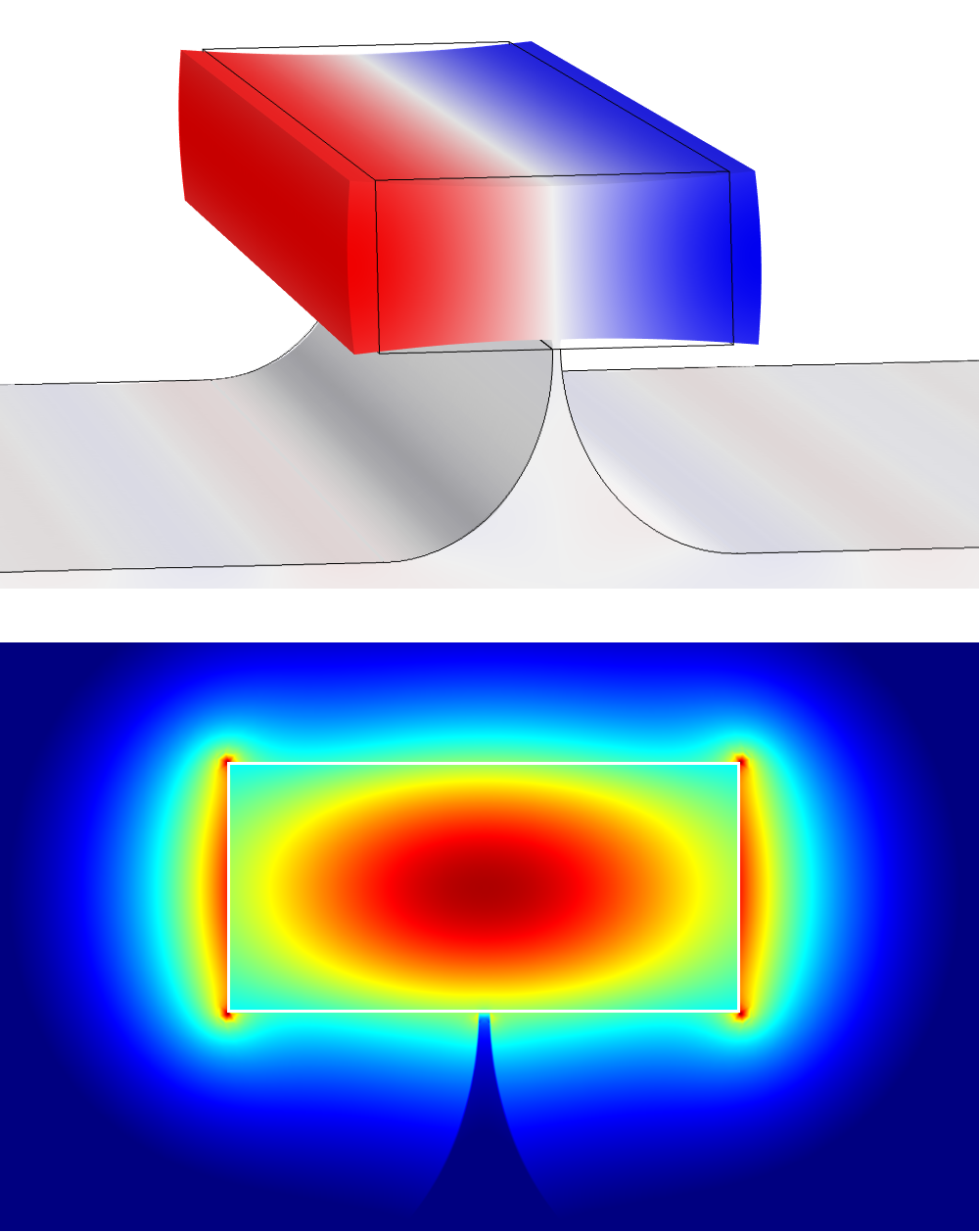}}};
    		\begin{scope}[x={(image.south east)},y={(image.north west)}]

		\node at (-0.04,0.96) {\large{\textbf{d}}};
		\node at (-0.04,0.45) {\large{\textbf{e}}};
		\node at (0.51,0.279) {\textcolor{white}{{$\bf{|E|}^2$}}};
		\node at (0.73,0.03) {\textcolor{white}{{$0$}}};
		\node at (0.98,0.03) {\textcolor{white}{{$1$}}};
		\node at (0.71,0.945) {\textcolor{black}{{$-1$}}};
		\node at (0.98,0.945) {\textcolor{black}{{$1$}}};

		\draw[->,thick,densely dotted,decorate,decoration={snake,amplitude=0.8mm,segment length=2.5mm,pre length=0.5mm,post length=2mm}] (0.65,0.64) -- (0.83,0.65);
		\node at (0.84,0.62) {{\large{{$\mathbf{\boldsymbol\tau}$}}}};
		
		\draw[->,thick,densely dotted,decorate,decoration={snake,amplitude=0.8mm,segment length=2.5mm,pre length=0.5mm,post length=2mm}] (0.35,0.63) -- (0.17,0.62);
		\node at (0.16,0.59) {{\large{{$\mathbf{\boldsymbol\tau}$}}}};


		\draw[black!10,fill,opacity=0.3,rotate around={1.5:(0.45,0.785)}] (0.45,0.785) ellipse (2.5mm and 3.2mm);
		\draw[<-*,thick,rotate around={1.5:(0.47,0.785)}] (0.364,0.785) -- (0.47,0.785);
		
		\draw[black!10,fill,opacity=0.3,rotate around={1.5:(0.675,0.792)}] (0.675,0.792) ellipse (2.6mm and 3.2mm);
		\draw[*->,black,thick,rotate around={1.5:(0.655,0.792)}] (0.655,0.792) -- (0.765,0.792);
		
		\draw[->,gray,thick,rotate around={1.5:(0.355,0.783)}] (0.355,0.783) -- (0.29,0.783);
		\draw[black,black!10,opacity=0.5] (0.356,0.853) .. controls (0.353,0.79) .. (0.360,0.714);

		\draw[->,gray,thick,rotate around={1.5:(0.255,0.846)}] (0.255,0.846) -- (0.19,0.846);
		\draw[black,black!10,opacity=0.5] (0.256,0.915) .. controls (0.253,0.85) .. (0.260,0.785);

		\draw[->,gray,thick,rotate around={1.5:(0.155,0.883)}] (0.185,0.895) -- (0.12,0.895);
		\draw[black,black!10,opacity=0.5] (0.185,0.959) .. controls (0.182,0.89) .. (0.189,0.837);

		\draw[->,gray,thick,rotate around={1.5:(0.775,0.794)}] (0.775,0.794) --  (0.840,0.794);
		\draw[black,black!10,opacity=0.5] (0.769,0.862) .. controls (0.777,0.795) .. (0.773,0.721);


    		\end{scope}

		\begin{scope}[shift={(3.95,0.1)},rotate=0]
		\begin{axis}[
    		hide axis,
   		scale only axis,
    		height=0pt,
    		width=0pt,
    		colormap/jet,
		colorbar horizontal,
    		point meta min=0,
    		point meta max=1,
    		colorbar style={width=1cm,height=0.2cm,ticks=none,
at={(5,0.5)},anchor=south west}]
    		\addplot [draw=none] coordinates {(0,0)};
		\end{axis}
		\end{scope}
		\begin{scope}[x={(image.south east)},y={(image.north west)}]
		\draw[white] (0.76,0.016) rectangle (0.951,0.0458);
    		\end{scope}
		
		\pgfplotsset{
		colormap={wavelight}{
		color(0cm)=(red); color(1cm)=(white); color(2cm)=(blue)}
		}
		\begin{scope}[shift={(3.95,6.1)},rotate=0]
		\begin{axis}[
    		hide axis,
   		scale only axis,
    		height=0pt,
    		width=0pt,
    		colormap name = {wavelight},
		colorbar horizontal,
    		point meta min=0,
    		point meta max=1,
    		colorbar style={width=1cm,height=0.2cm,ticks=none,
at={(5,0.5)},anchor=south west}]
    		\addplot [draw=none] coordinates {(0,0)};
		\end{axis}
		\end{scope}
		\end{tikzpicture}
}
\caption{\textbf{A silicon wire on a pillar as an acoustic phonon cavity.} \textbf{a}, Top view of the silicon wire. Light propagates along the wire. It confines photons owing to the high optical contrast with the silicon dioxide substrate and the air. \textbf{b}, Unlike the photons, the phonons are trapped transversally. The leakage of phonons through the pillar determines their lifetime $\tau \approx 5 \, \text{ns}$. \textbf{c}, A scanning electron micrograph of the $450\times230 \, \text{nm}$ cross-section (see Methods). We can fabricate pillars as narrow as $15 \; \text{nm}$ reliably. \textbf{d}, The horizontal component of the observed hypersonic mode $\mathbf{u}$ (red: $-$, blue: $+$) aligns with the bulk electrostrictive (black arrows) and the boundary radiation pressure forces (grey arrows). \textbf{e}, Electric field norm of the quasi-TE optical mode.}
\vspace{-3mm}
\end{figure*}

The observed mechanical mode strongly interacts with the fundamental quasi-TE optical mode (fig.1e). The main contribution to the coupling stems from the good overlap between the horizontal optical forces and displacement profile. In particular, the bulk electrostrictive forces $ \mathbf{f_{\text{es}}}$ and the boundary radiation pressure  $\mathbf{f_{\text{rp}}}$ both point in the same direction as the displacement field $\mathbf{u}$ (fig.1d). Therefore they interfere constructively, leading to a total overlap $\langle \mathbf{f}, \mathbf{u}\rangle = \langle \mathbf{f_{\text{es}}}, \mathbf{u}\rangle +  \langle \mathbf{f_{\text{rp}}}, \mathbf{u}\rangle$ up to twice as large as each individual component. Since the SBS gain $G_{\text{SBS}}(\Omega_{\text{m}})$ at the phonon resonance frequency $\Omega_{\text{m}}$ scales as $|\langle \mathbf{f}, \mathbf{u}\rangle|^{2}$, the total scattering from pump to Stokes photons may be up to four times as efficient as by electrostriction or radiation pressure individually.

This view of force interference \cite{Rakich2012,Qiu2013} was confirmed for the first time in a hybrid silicon nitride/silicon waveguide \cite{Shin2013b}. In that case, the photons were confined to the silicon core but the phonons mostly to the silicon nitride membrane. In our work, both the photons and the phonons are confined to the same silicon core. The elastic mode (fig.1d) can be understood as the fundamental mode of a Fabry-P\'{e}rot cavity for hypersonic waves (fig.1b), formed by the silicon-air boundaries. Therefore, its frequency can be estimated as $\frac{\Omega_{\text{m}}}{2\pi} = \frac{v}{2w} = 9.4 \, \text{GHz}$ with $v = 8433 \, \text{m/s}$ the longitudinal speed of sound in silicon and $w = 450 \, \text{nm}$ the waveguide width.

To create the pillar structure, we start from a silicon-on-insulator wire fabricated by deep UV lithography\cite{Bogaerts2005} through the multi-project-wafer service ePIXfab (\url{www.ePIXfab.eu}). Next, we perform an additional oxide etch with diluted hydrofluoric acid. By carefully controlling the etching speed, a narrow pillar is left underneath the wire (fig.1a-c). Through this simple fabrication method, we obtain wires up to $4 \, \text{cm}$ long. To retain compactness, wires longer than $3 \, \text{mm}$ are coiled up into a low-footprint spiral. Despite the additional etch, the wires still exhibit optical propagation losses $\alpha$ as low as $2.6 \, \text{dB/cm}$.

In our experiments (fig.2), we investigate straight and spiral waveguides with lengths $L$ ranging from $1.4 \, \text{mm}$ to $4 \, \text{cm}$. We couple $1550 \; \text{nm}$ TE-light to the waveguides through focusing grating couplers\cite{Laere2007} and perform both gain (fig.2a-b) and cross-phase modulation (fig.2c-d) experiments. The resonances (fig.2a and c) observed in these experiments allow for a characterization of the photon-phonon coupling in two independent ways.

\begin{figure*}
\centering
\label{fig:2}
\subfigure{
%
%
%
%
\begin{tikzpicture}

\begin{axis}[scale=0.28*\textwidth/(4.5in),
width=4.52083333333333in,
height=3.565625in,
xmin=9.1,
xmax=9.3,
xlabel={Frequency spacing (GHz)},
xlabel style = {yshift=0ex},
ymin=0.2,
ymax=3.5,
ylabel={Stokes power (a.u.)},
ylabel style = {yshift=-3ex},
axis x line*=bottom,
axis y line*=left,
legend style={fill=none,draw=none,legend cell align=left,at={(0.5,-0.41)},anchor=north, legend columns=-1},
xtick = {9.15,9.2,9.25},
ytick = {1,2,3},
minor xtick={9.125,9.15,...,9.275},
minor ytick={1,1.5,...,3}
]
\addplot [
color=black,
dashed,
line width=1.0pt,
forget plot
]
table[row sep=crcr]{
9.1 2.80\\
9.2 2.80\\
};
\addplot [
color=blue,
line width=0.5pt,
mark size=2.0pt,
only marks,
mark=o,
mark options={solid},
forget plot
]
table[row sep=crcr]{
9 1.0017146907957\\
9.005 0.999953821001468\\
9.01 0.998175418771247\\
9.015 0.999321853734277\\
9.02 1.00102870159344\\
9.025 1.00033851002082\\
9.03 0.999442430130025\\
9.03500000000001 0.999646678240361\\
9.04000000000001 0.998480238730132\\
9.04500000000001 1.00135946378231\\
9.05000000000001 1.00045036850147\\
9.05500000000001 1.00107374922282\\
9.06000000000001 1.00685456733397\\
9.06500000000001 1.00888140270391\\
9.07000000000001 1.0143511450261\\
9.07500000000001 1.01245372189761\\
9.08000000000001 1.02156316350458\\
9.08500000000001 1.02859900333248\\
9.09000000000001 1.03318754603694\\
9.09500000000001 1.03783349319491\\
9.10000000000002 1.03906442709838\\
9.10500000000002 1.05189089510584\\
9.11000000000002 1.06365199953505\\
9.11500000000002 1.07398457474507\\
9.12000000000002 1.08727098555897\\
9.12500000000002 1.11012338580562\\
9.13000000000002 1.14358486717903\\
9.13500000000002 1.18823588277899\\
9.14000000000002 1.24231369824197\\
9.14500000000002 1.31300420399785\\
9.15000000000002 1.37616525476536\\
9.15500000000002 1.40663892247415\\
9.16000000000003 1.40253521894345\\
9.16500000000003 1.43013360380704\\
9.17000000000003 1.53902994033308\\
9.17500000000003 1.72034148646235\\
9.18000000000003 1.99024051312784\\
9.18500000000003 2.28209727473456\\
9.19000000000003 2.59330938272091\\
9.19500000000003 2.80726934885256\\
9.20000000000003 2.80300633924873\\
9.20500000000003 2.68738454851521\\
9.21000000000003 2.49618928112323\\
9.21500000000003 2.2206339429353\\
9.22000000000003 1.83484541952413\\
9.22500000000004 1.51720879992256\\
9.23000000000004 1.29186867900895\\
9.23500000000004 1.16771852814886\\
9.24000000000004 1.10492492267265\\
9.24500000000004 1.06450698125194\\
9.25000000000004 1.03929058744956\\
9.25500000000004 1.02143905737644\\
9.26000000000004 1.01076427645321\\
9.26500000000004 1.00440840776988\\
9.27000000000004 0.999973248844446\\
9.27500000000004 0.999181744519633\\
9.28000000000004 0.997919176795742\\
9.28500000000004 0.998513733500502\\
9.29000000000005 0.998129922618092\\
9.29500000000005 0.998944742404272\\
9.30000000000005 1.00021169469638\\
9.30500000000005 1.00056734784903\\
9.31000000000005 1.00223826772553\\
9.31500000000005 1.00293298222434\\
};

\addplot [
color=red,
solid,
line width=2.0pt,
forget plot
]
table[row sep=crcr]{
9 1.01354692810236\\
9.001 1.01368364940052\\
9.002 1.01382244676675\\
9.003 1.01396336236004\\
9.004 1.01410643941271\\
9.005 1.0142517222633\\
9.006 1.01439925639062\\
9.007 1.01454908844908\\
9.008 1.01470126630524\\
9.009 1.01485583907569\\
9.01 1.01501285716638\\
9.011 1.01517237231334\\
9.012 1.01533443762489\\
9.013 1.0154991076255\\
9.014 1.0156664383012\\
9.015 1.01583648714673\\
9.016 1.01600931321449\\
9.017 1.01618497716532\\
9.018 1.0163635413212\\
9.019 1.01654506972006\\
9.02 1.01672962817256\\
9.021 1.01691728432122\\
9.022 1.01710810770173\\
9.023 1.01730216980677\\
9.024 1.01749954415224\\
9.025 1.01770030634622\\
9.026 1.01790453416059\\
9.027 1.01811230760556\\
9.028 1.01832370900723\\
9.029 1.0185388230882\\
9.03 1.01875773705159\\
9.031 1.01898054066841\\
9.032 1.01920732636856\\
9.033 1.01943818933561\\
9.034 1.01967322760552\\
9.035 1.01991254216952\\
9.036 1.02015623708132\\
9.037 1.02040441956882\\
9.038 1.02065720015071\\
9.039 1.02091469275795\\
9.04 1.02117701486056\\
9.041 1.02144428759993\\
9.042 1.02171663592687\\
9.043 1.0219941887458\\
9.044 1.02227707906522\\
9.045 1.02256544415502\\
9.046 1.0228594257107\\
9.047 1.0231591700251\\
9.048 1.02346482816784\\
9.049 1.02377655617304\\
9.05 1.0240945152356\\
9.051 1.02441887191655\\
9.052 1.02474979835803\\
9.053 1.02508747250828\\
9.054 1.02543207835724\\
9.055 1.02578380618334\\
9.056 1.02614285281214\\
9.057 1.02650942188725\\
9.058 1.02688372415456\\
9.059 1.02726597776018\\
9.06 1.02765640856303\\
9.061 1.02805525046292\\
9.062 1.02846274574487\\
9.063 1.02887914544071\\
9.064 1.02930470970894\\
9.065 1.0297397082338\\
9.066 1.0301844206448\\
9.067 1.03063913695781\\
9.068 1.03110415803908\\
9.069 1.03157979609342\\
9.07 1.03206637517806\\
9.071 1.0325642317438\\
9.072 1.03307371520494\\
9.073 1.0335951885399\\
9.074 1.03412902892432\\
9.075 1.03467562839877\\
9.076 1.03523539457299\\
9.077 1.03580875136931\\
9.078 1.03639613980737\\
9.079 1.03699801883304\\
9.08 1.03761486619425\\
9.081 1.03824717936676\\
9.082 1.03889547653326\\
9.083 1.03956029761913\\
9.084 1.04024220538873\\
9.085 1.04094178660619\\
9.086 1.04165965326518\\
9.087 1.04239644389208\\
9.088 1.04315282492789\\
9.089 1.04392949219396\\
9.09 1.04472717244769\\
9.091 1.04554662503422\\
9.092 1.04638864364108\\
9.093 1.04725405816299\\
9.094 1.04814373668476\\
9.095 1.04905858759088\\
9.096 1.04999956181087\\
9.097 1.05096765521058\\
9.098 1.05196391114003\\
9.099 1.0529894231497\\
9.1 1.05404533788774\\
9.101 1.05513285819202\\
9.102 1.05625324639185\\
9.103 1.05740782783554\\
9.104 1.05859799466155\\
9.105 1.05982520983212\\
9.106 1.06109101145037\\
9.107 1.06239701738354\\
9.108 1.06374493021682\\
9.109 1.06513654256492\\
9.11 1.06657374277061\\
9.111 1.06805852102212\\
9.112 1.06959297592459\\
9.113 1.07117932156365\\
9.114 1.07281989510282\\
9.115 1.07451716496072\\
9.116 1.07627373961798\\
9.117 1.0780923771087\\
9.118 1.07997599525685\\
9.119 1.08192768272347\\
9.12 1.08395071093729\\
9.121 1.08604854698837\\
9.122 1.08822486757257\\
9.123 1.09048357408301\\
9.124 1.09282880895487\\
9.125 1.09526497338036\\
9.126 1.09779674652277\\
9.127 1.10042910637173\\
9.128 1.1031673523966\\
9.129 1.10601713017119\\
9.13 1.10898445816111\\
9.131 1.11207575688511\\
9.132 1.1152978806845\\
9.133 1.11865815235878\\
9.134 1.12216440095375\\
9.135 1.12582500301853\\
9.136 1.12964892768162\\
9.137 1.13364578593363\\
9.138 1.13782588454527\\
9.139 1.14220028509519\\
9.14 1.14678086863187\\
9.141 1.15158040654908\\
9.142 1.15661263831486\\
9.143 1.16189235675923\\
9.144 1.16743550169773\\
9.145 1.17325926274433\\
9.146 1.17938219224978\\
9.147 1.1858243293884\\
9.148 1.19260733650749\\
9.149 1.19975464894629\\
9.15 1.20729163962519\\
9.151 1.21524579979496\\
9.152 1.22364693741731\\
9.153 1.23252739471317\\
9.154 1.24192228645595\\
9.155 1.25186976058955\\
9.156 1.26241128269651\\
9.157 1.27359194570937\\
9.158 1.28546080601328\\
9.159 1.29807124669003\\
9.16 1.31148136804792\\
9.161 1.3257544046947\\
9.162 1.34095916715055\\
9.163 1.357170504241\\
9.164 1.37446978010282\\
9.165 1.39294535638129\\
9.166 1.41269306584479\\
9.167 1.43381665788693\\
9.168 1.45642818884877\\
9.169 1.48064832032112\\
9.17 1.5066064760571\\
9.171 1.53444079224597\\
9.172 1.56429777605412\\
9.173 1.59633156293783\\
9.174 1.63070263384093\\
9.175 1.6675758189029\\
9.176 1.70711737524806\\
9.177 1.74949088440831\\
9.178 1.79485167322682\\
9.179 1.84333942647286\\
9.18 1.89506863913057\\
9.181 1.95011656520441\\
9.182 2.00850837715118\\
9.183 2.0701993806408\\
9.184 2.13505436271234\\
9.185 2.20282451784788\\
9.186 2.27312291969805\\
9.187 2.34540019080039\\
9.188 2.41892283677566\\
9.189 2.49275756659857\\
9.19 2.56576565428269\\
9.191 2.63661177033071\\
9.192 2.70379143185418\\
9.193 2.76568000920891\\
9.194 2.82060392433743\\
9.195 2.86693136853605\\
9.196 2.90317598852642\\
9.197 2.92810332517026\\
9.198 2.9408273337425\\
9.199 2.94088398148771\\
9.2 2.92827118465802\\
9.201 2.9034489899753\\
9.202 2.86729997241428\\
9.203 2.82105589483816\\
9.204 2.76620133557924\\
9.205 2.70436727978456\\
9.206 2.63722735904946\\
9.207 2.56640698366582\\
9.208 2.49341195350967\\
9.209 2.41957925272902\\
9.21 2.34604941837791\\
9.211 2.27375756072177\\
9.212 2.2034388926051\\
9.213 2.13564433869188\\
9.214 2.07076216393191\\
9.215 2.00904229229611\\
9.216 1.95062084169624\\
9.217 1.89554321599775\\
9.218 1.84378478100385\\
9.219 1.79526867595386\\
9.22 1.74988067983977\\
9.221 1.70748128626588\\
9.222 1.66791527194253\\
9.223 1.63101910170813\\
9.224 1.5966265221925\\
9.225 1.56457267615385\\
9.226 1.53469703399625\\
9.227 1.50684539729543\\
9.228 1.48087118712428\\
9.229 1.45663619088117\\
9.23 1.43401090679358\\
9.231 1.41287459582958\\
9.232 1.39311512630836\\
9.233 1.37462867661606\\
9.234 1.35731934552306\\
9.235 1.34109870704037\\
9.236 1.32588533695893\\
9.237 1.31160433066086\\
9.238 1.29818682602024\\
9.239 1.28556954084892\\
9.24 1.27369433107942\\
9.241 1.26250777346281\\
9.242 1.25196077479799\\
9.243 1.24200820844478\\
9.244 1.23260857798268\\
9.245 1.22372370726947\\
9.246 1.21531845575452\\
9.247 1.20736045765551\\
9.248 1.19981988347362\\
9.249 1.19266922226777\\
9.25 1.18588308311064\\
9.251 1.1794380141893\\
9.252 1.17331233807861\\
9.253 1.16748600179677\\
9.254 1.1619404403418\\
9.255 1.15665845250067\\
9.256 1.15162408781657\\
9.257 1.14682254369021\\
9.258 1.14224007167851\\
9.259 1.13786389213631\\
9.26 1.13368211642344\\
9.261 1.12968367597125\\
9.262 1.12585825756807\\
9.263 1.12219624428382\\
9.264 1.11868866150861\\
9.265 1.11532712763085\\
9.266 1.11210380892555\\
9.267 1.10901137826496\\
9.268 1.10604297730105\\
9.269 1.10319218180312\\
9.27 1.10045296986421\\
9.271 1.09781969271753\\
9.272 1.09528704792899\\
9.273 1.09285005475416\\
9.274 1.09050403146815\\
9.275 1.08824457449504\\
9.276 1.08606753917992\\
9.277 1.08396902206118\\
9.278 1.08194534451408\\
9.279 1.07999303764857\\
9.28 1.07810882835515\\
9.281 1.07628962640222\\
9.282 1.07453251249729\\
9.283 1.07283472723226\\
9.284 1.07119366084011\\
9.285 1.06960684369709\\
9.286 1.06807193750996\\
9.287 1.06658672713346\\
9.288 1.06514911296791\\
9.289 1.06375710389108\\
9.29 1.0624088106826\\
9.291 1.06110243990258\\
9.292 1.05983628818959\\
9.293 1.05860873694575\\
9.294 1.05741824737979\\
9.295 1.05626335588106\\
9.296 1.0551426696998\\
9.297 1.05405486291102\\
9.298 1.05299867264118\\
9.299 1.05197289553851\\
9.3 1.05097638446934\\
9.301 1.05000804542431\\
9.302 1.04906683461942\\
9.303 1.04815175577824\\
9.304 1.04726185758259\\
9.305 1.04639623127995\\
9.306 1.04555400843679\\
9.307 1.04473435882788\\
9.308 1.04393648845226\\
9.309 1.04315963766742\\
9.31 1.04240307943375\\
9.311 1.0416661176619\\
9.312 1.04094808565636\\
9.313 1.0402483446488\\
9.314 1.03956628241562\\
9.315 1.03890131197394\\
};

\end{axis}
%
%
%
%

\begin{axis}[xshift=2.8cm,
yshift=2.05cm,
scale=0.12*\textwidth/(4.52in),
width=4.3083333333333in,
height=3.565625in,
xmin=9.1,
xmax=9.3,
xlabel={},
ymin=0.2,
ymax=1.1,
ylabel={},
ylabel style = {yshift=-2ex},
axis x line*=top,
axis y line*=right,
xtick = {9.15,9.25},
ytick = {0.35,1},
minor xtick={9.15,9.2,9.25},
minor ytick={0.35,0.675,1.0},
legend style={draw=black,fill=white,legend cell align=left}
]
\addplot [
color=black,
dashed,
line width=0.5pt,
forget plot
]
table[row sep=crcr]{
9.2 0.35\\
9.3 0.35\\
};
\addplot [
color=blue,
line width=0.5pt,
mark size=2.0pt,
only marks,
mark=o,
mark options={solid},
forget plot
]
table[row sep=crcr]{
9.1 0.997161131669717\\
9.105 0.99934270234145\\
9.11 1.00321934934874\\
9.115 1.00110594034073\\
9.12 0.999874687840165\\
9.125 0.99770277851358\\
9.13 0.992965768489658\\
9.13500000000001 0.982111731220619\\
9.14000000000001 0.965671246522326\\
9.14500000000001 0.945448899167856\\
9.15000000000001 0.921833881921395\\
9.15500000000001 0.908278816412379\\
9.16000000000001 0.91250427279308\\
9.16500000000001 0.908482960045865\\
9.17000000000001 0.887446879643439\\
9.17500000000001 0.828954045007436\\
9.18000000000001 0.752585648938028\\
9.18500000000001 0.651531773505739\\
9.19000000000001 0.539386996464479\\
9.19500000000001 0.432715081752472\\
9.20000000000002 0.39874864982401\\
9.20500000000002 0.412069001451825\\
9.21000000000002 0.481575059788388\\
9.21500000000002 0.606295639866038\\
9.22000000000002 0.748191699036402\\
9.22500000000002 0.857697954105238\\
9.23000000000002 0.923811003261252\\
9.23500000000002 0.95658451208304\\
9.24000000000002 0.973597841888254\\
9.24500000000002 0.986371125374808\\
9.25000000000002 0.993666693233924\\
9.25500000000002 0.998385781622393\\
9.26000000000002 0.999977665152857\\
9.26500000000003 1.00137743519456\\
9.27000000000003 1.00225252737824\\
9.27500000000003 1.00484617227479\\
9.28000000000003 1.00261161862196\\
9.28500000000003 1.00111475313395\\
9.29000000000003 0.999325447901544\\
9.29500000000003 0.998302530629641\\
9.30000000000003 0.997432688692453\\
9.30500000000003 0.995909863841002\\
9.31000000000003 0.991584062073087\\
9.31500000000003 0.985801425008187\\
9.32000000000003 0.984761923598125\\
9.32500000000003 0.981115937421041\\
};
\addplot [
color=red,
solid,
line width=2.0pt,
forget plot
]
table[row sep=crcr]{
9.1 0.985891258771814\\
9.101 0.985612139809029\\
9.102 0.985324717363014\\
9.103 0.98502866135885\\
9.104 0.984723625291674\\
9.105 0.984409245244107\\
9.106 0.98408513883509\\
9.107 0.983750904094663\\
9.108 0.983406118258719\\
9.109 0.983050336477249\\
9.11 0.982683090428974\\
9.111 0.982303886834629\\
9.112 0.981912205860433\\
9.113 0.981507499402502\\
9.114 0.981089189242057\\
9.115 0.980656665060359\\
9.116 0.980209282301194\\
9.117 0.979746359867566\\
9.118 0.979267177637965\\
9.119 0.978770973786104\\
9.12 0.978256941886416\\
9.121 0.977724227785847\\
9.122 0.977171926220478\\
9.123 0.976599077153289\\
9.124 0.97600466180699\\
9.125 0.975387598363061\\
9.126 0.974746737295132\\
9.127 0.974080856301434\\
9.128 0.973388654797293\\
9.129 0.972668747924384\\
9.13 0.971919660028758\\
9.131 0.971139817554369\\
9.132 0.97032754129292\\
9.133 0.96948103792418\\
9.134 0.968598390773564\\
9.135 0.967677549705361\\
9.136 0.966716320060671\\
9.137 0.965712350538614\\
9.138 0.964663119907544\\
9.139 0.963565922419707\\
9.14 0.962417851787947\\
9.141 0.961215783566159\\
9.142 0.959956355756428\\
9.143 0.958635947444495\\
9.144 0.957250655241386\\
9.145 0.955796267282248\\
9.146 0.954268234503285\\
9.147 0.95266163888409\\
9.148 0.950971158304794\\
9.149 0.949191027625335\\
9.15 0.947314995547167\\
9.151 0.945336276765531\\
9.152 0.943247498862705\\
9.153 0.941040643329347\\
9.154 0.93870698003182\\
9.155 0.936236994368702\\
9.156 0.933620306280135\\
9.157 0.930845580190222\\
9.158 0.927900424877674\\
9.159 0.924771282186274\\
9.16 0.921443303409797\\
9.161 0.917900212122597\\
9.162 0.914124152188169\\
9.163 0.910095519678909\\
9.164 0.90579277750213\\
9.165 0.901192251680835\\
9.166 0.896267908525031\\
9.167 0.890991112408155\\
9.168 0.885330364614001\\
9.169 0.879251024851819\\
9.17 0.872715018698215\\
9.171 0.865680536614129\\
9.172 0.858101733569181\\
9.173 0.849928443033282\\
9.174 0.841105925623205\\
9.175 0.831574681598916\\
9.176 0.821270368413751\\
9.177 0.810123880494495\\
9.178 0.7980616693342\\
9.179 0.785006408805725\\
9.18 0.770878144129786\\
9.181 0.755596103337961\\
9.182 0.739081396218121\\
9.183 0.721260873981237\\
9.184 0.702072465329658\\
9.185 0.681472326545901\\
9.186 0.659444120164169\\
9.187 0.63601063170204\\
9.188 0.611247696810937\\
9.189 0.585299984947633\\
9.19 0.558397524528525\\
9.191 0.530870962072298\\
9.192 0.503162536298143\\
9.193 0.475828903255634\\
9.194 0.44953175038324\\
9.195 0.425013151734396\\
9.196 0.403055183873628\\
9.197 0.384427105120351\\
9.198 0.369827115441457\\
9.199 0.359827526629159\\
9.2 0.35483090428722\\
9.201 0.355041098060123\\
9.202 0.360449500334337\\
9.203 0.370835611258873\\
9.204 0.385781966584336\\
9.205 0.404704351345663\\
9.206 0.426896738616252\\
9.207 0.451586708842946\\
9.208 0.477993575284646\\
9.209 0.505380423590432\\
9.21 0.533093300269933\\
9.211 0.560584571248241\\
9.212 0.587421192297492\\
9.213 0.613281076386595\\
9.214 0.63794164170886\\
9.215 0.661264352924023\\
9.216 0.683178194529131\\
9.217 0.703664005394927\\
9.218 0.722740727352869\\
9.219 0.740453979278333\\
9.22 0.756866959060329\\
9.221 0.772053451410891\\
9.222 0.786092622653546\\
9.223 0.799065265372091\\
9.224 0.811051180137406\\
9.225 0.8221274248723\\
9.226 0.832367210718005\\
9.227 0.841839269043624\\
9.228 0.850607554112321\\
9.229 0.85873117888886\\
9.23 0.866264507789182\\
9.231 0.873257350646132\\
9.232 0.879755217785377\\
9.233 0.885799607835331\\
9.234 0.89142830858641\\
9.235 0.896675697578719\\
9.236 0.901573033702021\\
9.237 0.906148734383894\\
9.238 0.910428635264616\\
9.239 0.914436230869486\\
9.24 0.918192895885559\\
9.241 0.921718087375576\\
9.242 0.925029528722549\\
9.243 0.928143376373141\\
9.244 0.931074370592498\\
9.245 0.933835971499451\\
9.246 0.936440481647883\\
9.247 0.938899156378563\\
9.248 0.941222303101161\\
9.249 0.9434193705883\\
9.25 0.945499029279566\\
9.251 0.947469243508495\\
9.252 0.949337336482422\\
9.253 0.951110048765685\\
9.254 0.95279359094258\\
9.255 0.954393691067701\\
9.256 0.955915637448336\\
9.257 0.957364317246452\\
9.258 0.958744251335969\\
9.259 0.960059625804523\\
9.26 0.961314320446999\\
9.261 0.962511934560791\\
9.262 0.963655810319314\\
9.263 0.964749053970436\\
9.264 0.96579455508001\\
9.265 0.966795004017051\\
9.266 0.967752907856042\\
9.267 0.968670604853189\\
9.268 0.969550277636858\\
9.269 0.970393965237538\\
9.27 0.97120357406964\\
9.271 0.971980887965689\\
9.272 0.972727577353034\\
9.273 0.973445207653986\\
9.274 0.974135246981983\\
9.275 0.97479907319906\\
9.276 0.975437980393292\\
9.277 0.976053184829058\\
9.278 0.976645830417705\\
9.279 0.977216993751531\\
9.28 0.977767688739804\\
9.281 0.97829887088181\\
9.282 0.97881144120852\\
9.283 0.979306249921513\\
9.284 0.979784099755043\\
9.285 0.980245749084725\\
9.286 0.980691914804144\\
9.287 0.981123274988722\\
9.288 0.981540471364397\\
9.289 0.981944111597102\\
9.29 0.982334771417565\\
9.291 0.98271299659469\\
9.292 0.983079304769582\\
9.293 0.983434187161217\\
9.294 0.98377811015383\\
9.295 0.984111516775196\\
9.296 0.984434828074205\\
9.297 0.984748444405427\\
9.298 0.985052746627702\\
9.299 0.985348097223199\\
9.3 0.985634841342889\\
9.301 0.985913307783831\\
9.302 0.986183809903287\\
9.303 0.986446646474241\\
9.304 0.986702102486536\\
9.305 0.986950449897524\\
9.306 0.987191948335796\\
9.307 0.987426845761291\\
9.308 0.987655379084826\\
9.309 0.987877774749844\\
9.31 0.988094249278985\\
9.311 0.988305009787865\\
9.312 0.988510254468272\\
9.313 0.988710173042844\\
9.314 0.988904947193105\\
9.315 0.989094750962633\\
9.316 0.989279751136974\\
9.317 0.989460107601827\\
9.318 0.989635973680886\\
9.319 0.989807496454648\\
9.32 0.9899748170614\\
9.321 0.990138070981499\\
9.322 0.990297388305995\\
9.323 0.990452893990569\\
9.324 0.990604708095697\\
9.325 0.99075294601387\\
};
\end{axis}
\node[draw=none,text width=0.9cm] at (0.56,2.42) {\footnotesize 4.4 dB};
\node[draw=none,text width=1cm] at (4.06,2.05) {\footnotesize -4.4 dB};
\draw[black,<->,dashed,thick] (1.75,1.5) -- (2.55,1.5);
\node[draw=none,text width=1.2cm] at (2.18,1.15) {\footnotesize 40 MHz};
\node[draw=none] at (1.0,0.3) {\footnotesize $L_{\text{eff}}=1.5 \,\text{cm}$};
\node[draw=none] at (3.5,0.29) {\footnotesize $P_{\text{p}} = 35 \, \text{mW}$};
\node at (-0.7,-0.7) {\large{\textbf{a}}};
\end{tikzpicture}%
}
\subfigure{
\hspace{2mm}
\beginpgfgraphicnamed{fig:SBSgain}
\begin{tikzpicture}[>=stealth']
\node at (0.25,-3.2) {\large{\textbf{b}}};
\draw[black,thick] (0,0) rectangle (1.5,0.8) ;
\draw[red,thick] (0.1,0.4) -- (1.2,0.4);
\filldraw[red](1.2,0.4) circle (0.7mm);
\foreach \n in {0,...,11} {\draw[red,rotate around={360/11*(\n-1):(1.2,0.4)}] (1.2,0.4) -- (1.45,0.4);};
\foreach \n in {0,...,11} {\draw[red,rotate around={(360/11*(\n-1)+360/22):(1.2,0.4)}] (1.2,0.4) -- (1.38,0.4);};
\node at (0.34,0.56) {{$\lambda_{P}$}};
\node at (0.75,-0.2) {1550 \hspace{-0.5mm}{nm}};

\draw[blue,very thick] (1.5,0.4) -- (2,0.4);
\draw[thick] (2,0.3) rectangle (2.2,0.5);
\draw[thick] (2,0.5) -- (2.2,0.3);
\node at (2.45,0.18) {\scriptsize{50\%}};

\draw[blue,very thick] (2.1,0.3) -- (2.1,-1.3);
\draw[blue,very thick] (2.08,-1.29) -- (2.6,-1.29);
\begin{scope}[rotate around={90:(3.0,-1.06)},shift={(0.6,1.12)}]
\draw[thick] (2.7,-1.16) circle (0.1);
\draw[thick] (3.0,-1.06) circle (0.2);
\draw[thick] (3.3,-1.16) circle (0.1);
\end{scope}
\begin{scope}[shift={(2.28,-0.48)}]
\node[rotate=90] at (0.0,0) {\scriptsize{FPC}};
\end{scope}

\begin{scope}[shift={(-1.3,0.0)}]
\node at (4.3,-1.3) {\scriptsize{IM}};
\draw[thick] (3.9,-1.6) rectangle (4.7,-1.0);
\draw[very thick,brown] (4.3,-0.7) -- (4.3,-0.99);
\draw[thick] (4.3,-0.5) circle (0.2);
\node at (4.3,-0.71) {\huge$\tilde{}$};
\node at (4.80,-0.21) {\scriptsize{10 \hspace{-0.5mm}{GHz}}};

\draw[blue,very thick] (4.7,-1.3) -- (5.2,-1.3);
\draw[thick] (5.5,-1.3) circle (0.3);
\draw[->,thick] (5.5,-1.3) ++(140:2.1mm) arc (-220:40:2.1mm) --++(120:1mm);
\draw[thick] (5.45,-0.73) -- (5.45,-0.67) -- (5.55,-0.67) -- (5.55,-0.73);
\node at (5.72,-0.64) {\tiny{LT}};
\draw[blue,very thick] (5.5,-0.99) -- (5.5,-0.7);
\draw[blue,very thick] (5.81,-1.3) -- (6.31,-1.3);

\draw[thick] (6.32,-1.5) -- (6.32,-1.1);
\draw[thick] (6.38,-1.5) -- (6.38,-1.1);
\draw[thick] (6.44,-1.5) -- (6.44,-1.1);
\draw[thick] (6.50,-1.5) -- (6.50,-1.1);
\draw[blue,very thick] (6.51,-1.3) -- (7.0,-1.3);
\node at (6.41,-0.95) {\scriptsize{FBG}};

\draw[thick] (7.0,-1.5) -- (7.0,-1.1) -- (7.4,-1.3) -- (7.0,-1.5) -- cycle;
\node at (7.2,-0.95) {\scriptsize{EDFA}};
\draw[blue, very thick] (7.38,-1.3) -- (7.91,-1.3);
\draw[thick] (7.91,-1.5) rectangle (8.41,-1.1);
\node at (8.16,-0.95) {\scriptsize{BPF}};
\draw[thick] (8.0,-1.45) .. controls (8.15,-1.05) .. (8.3,-1.45);
\end{scope}

\draw[blue, very thick] (7.11,-1.3) -- (7.63,-1.3);
\draw[blue, very thick] (7.61,-1.3) -- (7.61,0.3);
\draw[thick] (7.51,0.3) rectangle (7.71,0.5);
\draw[thick] (7.51,0.3) -- (7.71,0.5);

\begin{scope}[rotate around={-90:(3.0,-1.06)},shift={(-0.6,4.83)}]
\draw[thick] (2.7,-1.16) circle (0.1);
\draw[thick] (3.0,-1.06) circle (0.2);
\draw[thick] (3.3,-1.16) circle (0.1);
\end{scope}
\begin{scope}[shift={(7.45,-0.48)}]
\node[rotate=-90] at (0.0,0) {\scriptsize{FPC}};
\end{scope}

\draw[blue!40,fill,opacity=0.3,rounded corners] (3.84,-2.3) rectangle (5.4,-0.53);
\node[] at (4.65,-1.9) {\scriptsize{Remove $\lambda_{P}$}};
\node[] at (4.87,-2.15) {\scriptsize{and $\lambda_{AS}$}};

\draw[red!30,fill,opacity=0.2,rounded corners] (1.65,-2.35) rectangle (5.45,-0.03);
\node[] at (2.55,-2.1) {\scriptsize{Create Stokes}};

\draw[blue,very thick] (2.2,0.4) -- (3.1,0.4);
\draw[thick] (3.1,0.2) -- (3.1,0.6) -- (3.5,0.4) -- (3.1,0.2) -- cycle;
\draw[blue,very thick] (3.48,0.4) -- (4.4,0.4);
\node at (3.3,0.73) {\scriptsize{EDFA}};

\draw[thick] (4.4,0.2) rectangle (4.9,0.6);
\node at (4.65,0.73) {\scriptsize{BPF}};
\draw[thick] (4.5,0.25) .. controls (4.65,0.65) .. (4.8,0.25);

\draw[blue,very thick] (4.91,0.4) -- (7.51,0.4);
\draw[thick] (5.9,0.52) circle (0.1);
\draw[thick] (6.2,0.62) circle (0.2);
\draw[thick] (6.5,0.52) circle (0.1);
\node[] at (6.22,0.22) {\scriptsize{FPC}};
\node at (7.28,0.18) {\scriptsize{50\%}};

\draw[blue,very thick] (7.61,0.5) -- (7.61,0.75);
\draw[thick,fill,black!90] (7.51,0.75) .. controls (7.56,0.95) and (7.66,0.95) .. (7.71,0.75);
\node[] at (7.95,0.85) {\scriptsize{PD}};
\draw[blue,very thick] (7.71,0.4) -- (8.3,0.4);

\draw[thick] (8.6,0.4) circle (0.3);
\draw[->,thick] (8.6,0.15) ++(140:2.1mm) arc (220:-40:2.1mm) --++(-120:1mm);
\draw[blue,very thick] (8.6,0.1) -- (8.6,-0.2);
\draw[thick] (8.55,-0.17) -- (8.55,-0.22) -- (8.65,-0.22) -- (8.65,-0.17);
\node at (8.52,-0.33) {\tiny{LT}};

\draw[blue,very thick] (8.9,0.4) -- (9.4,0.4);

 \node[anchor=south west,inner sep=0,rotate=0] (image) at (9.2,-0.95) {\raisebox{0pt}{\includegraphics[scale=0.06]{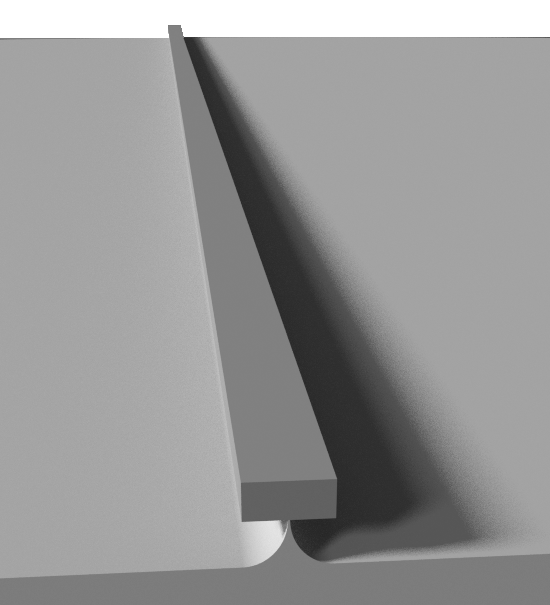}}};

\draw[black!30,fill,opacity=0.2,rounded corners] (9.1,-1.08) rectangle (10.4,0.14);
\draw[white,->,very thin,decorate,decoration={snake,amplitude=.2mm,segment length=1mm,post length=1mm,pre length=1mm}] (9.50,-0.1) -- (9.64,-0.65);
\begin{scope}[shift={(10.24,-0.5)}]
\node[rotate=-90] at (0.0,0) {\scriptsize{Chip}};
\end{scope}

\draw[rotate around={2.5:(9.483,0.06)}] (9.418,0.10) -- (9.544,0.10) -- (9.481,-0.005) -- (9.418,0.10) -- cycle;
\draw[very thin] (9.44,0.05) arc[start angle=150,end angle=40,x radius=0.5mm,y radius = 0.5mm];
\draw[very thin] (9.46,0.035) arc[start angle=150,end angle=40,x radius=0.25mm,y radius = 0.25mm];
\draw[thin] (9.4,0.4) .. controls (9.65,0.4) and (9.45,0.25) .. (9.48,0.12);

\begin{scope}[rotate around={180:(9.483,0.06)},shift={(-0.19,1.05)}]
\draw[rotate around={2.5:(9.483,0.06)}] (9.418,0.10) -- (9.544,0.10) -- (9.481,-0.005) -- (9.418,0.10) -- cycle;
\draw[very thin] (9.44,0.05) arc[start angle=150,end angle=40,x radius=0.5mm,y radius = 0.5mm];
\draw[very thin] (9.46,0.035) arc[start angle=150,end angle=40,x radius=0.25mm,y radius = 0.25mm];
\end{scope}

\draw[blue,very thick] (9.4,-1.3) -- (9.4,-1.9);
\draw[thin] (9.68,-1.05) .. controls (9.695,-1.25) and (9.38,-1.15) .. (9.4,-1.3);
\draw[->,thin] (9.6,0.3) .. controls (9.65,0.15) .. (9.6,-0.1);
\node at (10.03,0.22) {\tiny{{$\eta$$\approx$25\%}}};
\begin{scope}[shift={(-0.07,-1)}]
\draw[->,thin] (9.6,0.3) .. controls (9.55,0.15) .. (9.6,-0.1);
\end{scope}

\draw[thick] (9.3,-1.9) rectangle (9.5,-2.1);
\draw[thick] (9.3,-2.1) -- (9.5,-1.9);
\draw[blue,very thick] (9.3,-2.0) -- (8.3,-2.0);
\node at (8.8,-1.85) {\scriptsize{1\%}};
\begin{scope}[rotate around = {90:(7.51,0.75)},shift={(-2.85,-0.80)}]
\draw[thick,fill,black!90] (7.51,0.75) .. controls (7.56,0.95) and (7.66,0.95) .. (7.71,0.75);
\end{scope}
\node[] at (7.9,-2.0) {\scriptsize{PD}};

\draw[blue,very thick] (9.4,-2.1) -- (9.4,-2.85) -- (8.9,-2.85);
\node at (9.1,-2.4) {\scriptsize{99\%}};

\begin{scope}[rotate around={180:(5.8,-1.3)},shift={(-2.15,1.55)}]
\draw[blue,very thick] (4.7,-1.3) -- (5.2,-1.3);
\draw[thick] (5.5,-1.3) circle (0.3);
\draw[->,thick] (5.5,-1.3) ++(140:2.1mm) arc (-220:40:2.1mm) --++(120:1mm);
\draw[thick] (5.45,-0.73) -- (5.45,-0.67) -- (5.55,-0.67) -- (5.55,-0.73);
\node at (5.72,-0.74) {\tiny{LT}};
\draw[blue,very thick] (5.5,-0.99) -- (5.5,-0.7);
\draw[blue,very thick] (5.81,-1.3) -- (6.31,-1.3);

\draw[thick] (6.32,-1.5) -- (6.32,-1.1);
\draw[thick] (6.38,-1.5) -- (6.38,-1.1);
\draw[thick] (6.44,-1.5) -- (6.44,-1.1);
\draw[thick] (6.50,-1.5) -- (6.50,-1.1);
\draw[blue,very thick] (6.51,-1.3) -- (7.0,-1.3);
\node at (6.41,-0.95) {\scriptsize{FBG}};
\end{scope}
\begin{scope}[rotate around = {90:(7.51,0.75)},shift={(-3.7,0.7)}]
\draw[thick,fill,black!90] (7.51,0.75) .. controls (7.56,0.95) and (7.66,0.95) .. (7.71,0.75);
\end{scope}
\node[] at (6.4,-2.85) {\scriptsize{PD}};

\draw[blue!40,fill,opacity=0.2,rounded corners] (7.0,-3.55) rectangle (8.7,-2.25);
\node at (7.91,-2.4) {\scriptsize{Remove $\lambda_{P}$}};

\end{tikzpicture}
\endpgfgraphicnamed
}
\\
\subfigure{
\hspace{-1mm}
%
%
%
%
\begin{tikzpicture}

\begin{axis}[scale = 0.28*\textwidth/(4.5in),
width=4.52083333333333in,
height=3.565625in,
xmin=8.4,
xmax=10,
xlabel={Frequency spacing (GHz)},
ymin=0,
ymax=3.5,
ylabel={Sideband power (a.u.)},
ylabel style = {yshift=-3ex},
axis x line*=bottom,
axis y line*=left,
legend style={fill=none,draw=none,legend cell align=left,at={(0.5,1.2)},anchor=north, legend columns=-1},
xtick = {8.8,9.2,9.6},
ytick = {1,2,3},
minor xtick={8.6,8.8,...,9.8},
minor ytick={0.5,1,...,3}
]
\addplot [
color=blue,
line width=0.3pt,
mark size=2.0pt,
only marks,
mark=o,
mark options={solid},
]
table[row sep=crcr]{
8.5 1\\
8.51 0.963829023623971\\
8.52 0.993116048420934\\
8.53 0.931107875467831\\
8.54 0.933254300796991\\
8.55 0.972747223776966\\
8.56 0.972747223776966\\
8.57 0.941889596522842\\
8.58 0.948418463300898\\
8.59 0.959400631515934\\
8.6 0.988553094656939\\
8.61 0.946237161365793\\
8.62 0.979489985408699\\
8.63 0.928966386779937\\
8.64 0.96827785626125\\
8.65 0.922571427154763\\
8.66 0.963829023623971\\
8.67 0.990831944892768\\
8.68 0.984011105761134\\
8.69 0.984011105761134\\
8.7 1.0471285480509\\
8.71 1.03514216667934\\
8.72 1.05438689639126\\
8.73 1.09395636627209\\
8.74 1.05438689639126\\
8.75 1.08642562361707\\
8.76 1.0447202192208\\
8.77 1.05681750921366\\
8.78 1.07151930523761\\
8.79 1.10662378397767\\
8.8 1.12460497396693\\
8.81 1.11173172728159\\
8.82 1.17219536554813\\
8.83 1.16144861384034\\
8.84 1.18576874816716\\
8.85 1.15080038894444\\
8.86 1.20781383510678\\
8.87 1.17760597352081\\
8.88 1.23310483322891\\
8.89 1.17489755493953\\
8.9 1.18850222743702\\
8.91 1.21618600064637\\
8.92 1.1297959146728\\
8.93 1.15080038894444\\
8.94 1.14815362149688\\
8.95 1.1561122421921\\
8.96 1.11429453359173\\
8.97 1.17219536554813\\
8.98 1.18576874816716\\
8.99 1.23594743344451\\
9 1.25602996369488\\
9.01 1.2050359403718\\
9.02 1.22461619926505\\
9.03 1.27057410520854\\
9.04 1.24165230759241\\
9.05 1.27643880881134\\
9.06 1.25314117494142\\
9.07 1.32739445772974\\
9.08 1.29419584144999\\
9.09 1.36458313658893\\
9.1 1.35831344658715\\
9.11 1.37404197501252\\
9.12 1.51008015416415\\
9.13 1.51008015416415\\
9.14 1.69824365246175\\
9.15 1.8113400926196\\
9.16 2.07491351745491\\
9.17 2.1183611352485\\
9.18 2.25423921215243\\
9.19 2.3120647901756\\
9.2 2.5822601906346\\
9.21 3.11888958409394\\
9.22 3.14774831410132\\
9.23 2.76057785622035\\
9.24 1.5346169827993\\
9.25 0.691830970918937\\
9.26 0.36897759857015\\
9.27 0.319153785510076\\
9.28 0.411149721104522\\
9.29 0.479733448636689\\
9.3 0.542000890401624\\
9.31 0.609536897240169\\
9.32 0.650129690343091\\
9.33 0.659173895244322\\
9.34 0.693425806016569\\
9.35 0.71121351365333\\
9.36 0.762079010025412\\
9.37 0.749894209332456\\
9.38 0.760326276940182\\
9.39 0.78162780458833\\
9.4 0.810961057853841\\
9.41 0.826037949577179\\
9.42 0.807235030248838\\
9.43 0.824138115013003\\
9.44 0.833681184619635\\
9.45 0.833681184619635\\
9.46 0.839459986519398\\
9.47 0.860993752184601\\
9.48 0.85703784523037\\
9.49 0.872971368388112\\
9.5 0.872971368388112\\
9.51 0.924698173938223\\
9.52 0.901571137605957\\
9.53 0.931107875467831\\
9.54 0.931107875467831\\
9.55 0.916220490122\\
9.56 0.914113241470251\\
9.57 0.928966386779937\\
9.58 0.924698173938223\\
9.59 0.931107875467831\\
9.6 0.916220490122\\
9.61 0.933254300796991\\
9.62 0.933254300796991\\
9.63 0.954992586021436\\
9.64 0.924698173938223\\
9.65 0.961612278383665\\
9.66 0.96827785626125\\
9.67 0.931107875467831\\
9.68 0.928966386779937\\
9.69 0.937562006925881\\
9.7 0.941889596522842\\
9.71 0.909913272632252\\
9.72 0.903649473722302\\
9.73 0.924698173938223\\
9.74 0.928966386779937\\
9.75 0.891250938133746\\
9.76 0.903649473722302\\
9.77 0.941889596522842\\
9.78 0.920449571753172\\
9.79 0.937562006925881\\
9.8 0.954992586021436\\
9.81 0.924698173938223\\
9.82 0.922571427154763\\
9.83 0.922571427154763\\
9.84 0.909913272632252\\
9.85 0.931107875467831\\
9.86 0.924698173938223\\
9.87 0.939723310564638\\
9.88 0.916220490122\\
9.89 0.941889596522842\\
9.9 0.963829023623971\\
9.91 0.935405674147552\\
9.92 0.961612278383665\\
9.93 0.950604793656282\\
9.94 0.97050996724549\\
9.95 0.972747223776966\\
9.96 0.959400631515934\\
9.97 1.0046157902784\\
9.98 0.974989637717387\\
9.99 1\\
};
\addlegendentry{Experiment \, \,};

\addplot [
color=red,
solid,
line width=2.0pt,
]
table[row sep=crcr]{
8.5 1.08084201560568\\
8.51 1.08182859811189\\
8.52 1.08284332222589\\
8.53 1.08388740817944\\
8.54 1.08496214771078\\
8.55 1.08606890937486\\
8.56 1.08720914433343\\
8.57 1.08838439267631\\
8.58 1.08959629033151\\
8.59 1.09084657662901\\
8.6 1.09213710259167\\
8.61 1.09346984003576\\
8.62 1.09484689157495\\
8.63 1.09627050163399\\
8.64 1.09774306859279\\
8.65 1.09926715819845\\
8.66 1.10084551840232\\
8.67 1.10248109580144\\
8.68 1.10417705389037\\
8.69 1.10593679335988\\
8.7 1.10776397471502\\
8.71 1.10966254352708\\
8.72 1.11163675868404\\
8.73 1.11369122406228\\
8.74 1.11583092411203\\
8.75 1.11806126393175\\
8.76 1.12038811450467\\
8.77 1.12281786388889\\
8.78 1.12535747529348\\
8.79 1.12801455314381\\
8.8 1.13079741844507\\
8.81 1.13371519500389\\
8.82 1.13677790837314\\
8.83 1.13999659975971\\
8.84 1.14338345759572\\
8.85 1.14695197004413\\
8.86 1.1507171024181\\
8.87 1.15469550437986\\
8.88 1.1589057528977\\
8.89 1.16336863834748\\
8.9 1.16810750293484\\
8.91 1.17314864290548\\
8.92 1.17852178896382\\
8.93 1.18426068315284\\
8.94 1.19040377546057\\
8.95 1.19699507002615\\
8.96 1.20408515960375\\
8.97 1.21173249872631\\
8.98 1.22000498197043\\
8.99 1.22898191555501\\
9 1.2387565006957\\
9.01 1.24943898936478\\
9.02 1.26116073290201\\
9.03 1.27407942971006\\
9.04 1.28838600308553\\
9.05 1.30431372458265\\
9.06 1.32215047498249\\
9.07 1.342255457379\\
9.08 1.36508233381172\\
9.09 1.39121179846577\\
9.1 1.42139828601394\\
9.11 1.45663829924008\\
9.12 1.49827254032986\\
9.13 1.54814211223152\\
9.14 1.6088331357573\\
9.15 1.68406860260687\\
9.16 1.7793470059927\\
9.17 1.90298323167205\\
9.18 2.06771463912319\\
9.19 2.29254526555847\\
9.2 2.60086674868184\\
9.21 2.99000013435869\\
9.22 3.26424179226193\\
9.23 2.74213831417652\\
9.24 1.43276574805686\\
9.25 0.607241199045503\\
9.26 0.349484412877113\\
9.27 0.317269448602372\\
9.28 0.353261240003327\\
9.29 0.405333745314883\\
9.3 0.457242005815591\\
9.31 0.504343364933977\\
9.32 0.545782394724247\\
9.33 0.581906762055931\\
9.34 0.613385425432852\\
9.35 0.640908205847778\\
9.36 0.665091520983196\\
9.37 0.686458002592087\\
9.38 0.705441539931917\\
9.39 0.722399701554646\\
9.4 0.737626936701215\\
9.41 0.751366353364531\\
9.42 0.763819548043875\\
9.43 0.775154570593996\\
9.44 0.785512294265498\\
9.45 0.79501148626958\\
9.46 0.803752843197033\\
9.47 0.811822210253886\\
9.48 0.819293159284855\\
9.49 0.8262290629747\\
9.5 0.832684772247538\\
9.51 0.838707980003527\\
9.52 0.84434033580966\\
9.53 0.849618361883877\\
9.54 0.854574209727398\\
9.55 0.85923628830333\\
9.56 0.863629788133088\\
9.57 0.86777712062829\\
9.58 0.871698288046171\\
9.59 0.875411196387628\\
9.6 0.878931921148902\\
9.61 0.88227493393931\\
9.62 0.885453296473385\\
9.63 0.888478827248441\\
9.64 0.891362245261063\\
9.65 0.894113294346599\\
9.66 0.896740851104718\\
9.67 0.899253018870666\\
9.68 0.90165720978195\\
9.69 0.903960216655054\\
9.7 0.90616827611172\\
9.71 0.908287124167629\\
9.72 0.910322045308812\\
9.73 0.91227791592543\\
9.74 0.914159242842871\\
9.75 0.915970197581673\\
9.76 0.917714646886838\\
9.77 0.919396179990557\\
9.78 0.921018133007786\\
9.79 0.922583610809417\\
9.8 0.924095506671336\\
9.81 0.925556519958161\\
9.82 0.926969172066665\\
9.83 0.928335820825008\\
9.84 0.929658673519105\\
9.85 0.93093979869612\\
9.86 0.932181136876679\\
9.87 0.933384510291494\\
9.88 0.934551631744302\\
9.89 0.935684112691066\\
9.9 0.936783470614985\\
9.91 0.937851135767764\\
9.92 0.938888457339678\\
9.93 0.93989670911402\\
9.94 0.940877094655445\\
9.95 0.941830752076337\\
9.96 0.94275875842068\\
9.97 0.943662133700682\\
9.98 0.944541844617819\\
9.99 0.945398807996648\\
};
\addlegendentry{Fit};
\end{axis}
\node[draw=none] at (1.0,0.3) {\footnotesize $L_{\text{eff}}=1.2 \,\text{cm}$};
\node at (-0.7,-0.7) {\large{\textbf{c}}};
\end{tikzpicture}%
}
\subfigure{
\hspace{11mm}
\beginpgfgraphicnamed{fig:SBSgain}
\begin{tikzpicture}[>=stealth']
\node at (0.25,-3.2) {\large{\textbf{d}}};
\begin{scope}[shift={(-0.00,0.0)}]
\draw[black,thick] (0,0) rectangle (1.5,0.8) ;
\draw[green!50!black,thick] (0.1,0.4) -- (1.2,0.4);
\filldraw[green](1.2,0.4) circle (0.7mm);
\foreach \n in {0,...,11} {\draw[green!50!black,rotate around={360/11*(\n-1):(1.2,0.4)}] (1.2,0.4) -- (1.45,0.4);};
\foreach \n in {0,...,11} {\draw[green!50!black,rotate around={(360/11*(\n-1)+360/22):(1.2,0.4)}] (1.2,0.4) -- (1.38,0.4);};
\node at (0.52,0.57) {{Pump}};
\node at (0.75,-0.2) {1537 \hspace{-0.5mm}{nm}};
\end{scope}

\begin{scope}[shift={(5.6,-1.7)}]
\draw[black,thick] (0,0) rectangle (1.5,0.8) ;
\draw[red,thick] (0.1,0.4) -- (1.2,0.4);
\filldraw[red](1.2,0.4) circle (0.7mm);
\foreach \n in {0,...,11} {\draw[red,rotate around={360/11*(\n-1):(1.2,0.4)}] (1.2,0.4) -- (1.45,0.4);};
\foreach \n in {0,...,11} {\draw[red,rotate around={(360/11*(\n-1)+360/22):(1.2,0.4)}] (1.2,0.4) -- (1.38,0.4);};
\node at (0.50,0.57) {{Probe}};
\node at (0.75,1.0) {1550 \hspace{-0.5mm}{nm}};
\end{scope}


\begin{scope}[rotate around={180:(4.3,-1.3)},shift={(+0.75,-1.7)}]
\draw[thick] (3.9,-1.6) rectangle (4.7,-1.0);
\draw[very thick,brown] (4.3,-0.7) -- (4.3,-0.99);
\draw[thick] (4.3,-0.5) circle (0.2);
\node at (4.3,-0.29) {\huge$\tilde{}$};
\node at (4.3,-0.15) {\scriptsize{10 \hspace{-0.5mm}{GHz}}};
\node at (4.3,-1.3) {\scriptsize{IM}};
\end{scope}
\draw[blue, very thick] (2.64,0.4) -- (3.14,0.4);

\begin{scope}[shift={(-3.85,0.0)}]
\draw[thick] (5.9,0.52) circle (0.1);
\draw[thick] (6.2,0.62) circle (0.2);
\draw[thick] (6.5,0.52) circle (0.1);
\node[] at (6.22,0.22) {\scriptsize{FPC}};
\end{scope}
\draw[blue, very thick] (1.49,0.4) -- (2.64,0.4);


\draw[blue, very thick] (7.11,-1.3) -- (7.63,-1.3);
\draw[blue, very thick] (7.61,-1.3) -- (7.61,0.3);
\draw[thick] (7.51,0.3) rectangle (7.71,0.5);
\draw[thick] (7.51,0.3) -- (7.71,0.5);

\begin{scope}[rotate around={-90:(3.0,-1.06)},shift={(-0.6,4.83)}]
\draw[thick] (2.7,-1.16) circle (0.1);
\draw[thick] (3.0,-1.06) circle (0.2);
\draw[thick] (3.3,-1.16) circle (0.1);
\end{scope}
\begin{scope}[shift={(7.45,-0.48)}]
\node[rotate=-90] at (0.0,0) {\scriptsize{FPC}};
\end{scope}

\begin{scope}[shift={(1.35,0.0)}]
\draw[blue,very thick] (2.6,0.4) -- (3.1,0.4);
\draw[thick] (3.1,0.2) -- (3.1,0.6) -- (3.5,0.4) -- (3.1,0.2) -- cycle;
\draw[blue,very thick] (3.48,0.4) -- (3.95,0.4);
\node at (3.3,0.73) {\scriptsize{EDFA}};
\end{scope}

\begin{scope}[shift={(0.9,0.0)}]
\draw[thick] (4.4,0.2) rectangle (4.9,0.6);
\node at (4.65,0.73) {\scriptsize{BPF}};
\draw[thick] (4.5,0.25) .. controls (4.65,0.65) .. (4.8,0.25);
\end{scope}

\draw[blue,very thick] (5.81,0.4) -- (7.51,0.4);
\begin{scope}[shift={(0.5,0.0)}]
\draw[thick] (5.9,0.52) circle (0.1);
\draw[thick] (6.2,0.62) circle (0.2);
\draw[thick] (6.5,0.52) circle (0.1);
\end{scope}
\node at (7.28,0.18) {\scriptsize{50\%}};

\draw[blue,very thick] (7.61,0.5) -- (7.61,0.75);
\draw[thick,fill,black!90] (7.51,0.75) .. controls (7.56,0.95) and (7.66,0.95) .. (7.71,0.75);
\node[] at (7.95,0.85) {\scriptsize{PD}};
\draw[blue,very thick] (7.71,0.4) -- (8.3,0.4);

\draw[thick] (8.6,0.4) circle (0.3);
\draw[->,thick] (8.6,0.15) ++(140:2.1mm) arc (220:-40:2.1mm) --++(-120:1mm);
\draw[blue,very thick] (8.6,0.1) -- (8.6,-0.2);
\draw[thick] (8.55,-0.17) -- (8.55,-0.22) -- (8.65,-0.22) -- (8.65,-0.17);
\node at (8.52,-0.33) {\tiny{LT}};

\draw[blue,very thick] (8.9,0.4) -- (9.4,0.4);

 \node[anchor=south west,inner sep=0,rotate=0] (image) at (9.2,-0.95) {\raisebox{0pt}{\includegraphics[scale=0.06]{img/waveguide_view4.png}}};

\draw[black!30,fill,opacity=0.2,rounded corners] (9.1,-1.08) rectangle (10.4,0.14);
\draw[white,->,very thin,decorate,decoration={snake,amplitude=.2mm,segment length=1mm,post length=1mm,pre length=1mm}] (9.50,-0.1) -- (9.64,-0.65);
\begin{scope}[shift={(10.24,-0.5)}]
\node[rotate=-90] at (0.0,0) {\scriptsize{Chip}};
\end{scope}

\draw[rotate around={2.5:(9.483,0.06)}] (9.418,0.10) -- (9.544,0.10) -- (9.481,-0.005) -- (9.418,0.10) -- cycle;
\draw[very thin] (9.44,0.05) arc[start angle=150,end angle=40,x radius=0.5mm,y radius = 0.5mm];
\draw[very thin] (9.46,0.035) arc[start angle=150,end angle=40,x radius=0.25mm,y radius = 0.25mm];
\draw[thin] (9.4,0.4) .. controls (9.65,0.4) and (9.45,0.25) .. (9.48,0.12);

\begin{scope}[rotate around={180:(9.483,0.06)},shift={(-0.19,1.05)}]
\draw[rotate around={2.5:(9.483,0.06)}] (9.418,0.10) -- (9.544,0.10) -- (9.481,-0.005) -- (9.418,0.10) -- cycle;
\draw[very thin] (9.44,0.05) arc[start angle=150,end angle=40,x radius=0.5mm,y radius = 0.5mm];
\draw[very thin] (9.46,0.035) arc[start angle=150,end angle=40,x radius=0.25mm,y radius = 0.25mm];
\end{scope}

\draw[blue,very thick] (9.4,-1.3) -- (9.4,-1.9);
\draw[thin] (9.68,-1.05) .. controls (9.695,-1.25) and (9.38,-1.15) .. (9.4,-1.3);
\draw[->,thin] (9.6,0.3) .. controls (9.65,0.15) .. (9.6,-0.1);
\node at (10.03,0.22) {\tiny{{$\eta$$\approx$25\%}}};
\begin{scope}[shift={(-0.07,-1)}]
\draw[->,thin] (9.6,0.3) .. controls (9.55,0.15) .. (9.6,-0.1);
\end{scope}

\draw[thick] (9.3,-1.9) rectangle (9.5,-2.1);
\draw[thick] (9.3,-2.1) -- (9.5,-1.9);
\draw[blue,very thick] (9.3,-2.0) -- (8.3,-2.0);
\node at (8.8,-1.85) {\scriptsize{1\%}};
\begin{scope}[rotate around = {90:(7.51,0.75)},shift={(-2.85,-0.80)}]
\draw[thick,fill,black!90] (7.51,0.75) .. controls (7.56,0.95) and (7.66,0.95) .. (7.71,0.75);
\end{scope}
\node[] at (7.9,-2.0) {\scriptsize{PD}};

\draw[blue,very thick] (9.4,-2.1) -- (9.4,-2.85) -- (8.4,-2.85);
\node at (9.1,-2.4) {\scriptsize{99\%}};

\begin{scope}[shift={(-1.35,0.0)}]
\begin{scope}[rotate around={180:(5.8,-1.3)},shift={(-2.15,1.55)}]
\draw[thick] (5.5,-1.3) circle (0.3);
\draw[->,thick] (5.5,-1.3) ++(140:2.1mm) arc (-220:40:2.1mm) --++(120:1mm);
\draw[thick] (5.45,-0.73) -- (5.45,-0.67) -- (5.55,-0.67) -- (5.55,-0.73);
\node at (5.72,-0.74) {\tiny{LT}};
\draw[blue,very thick] (5.5,-0.99) -- (5.5,-0.7);
\draw[blue,very thick] (5.81,-1.3) -- (6.31,-1.3);

\draw[thick] (6.32,-1.5) -- (6.32,-1.1);
\draw[thick] (6.38,-1.5) -- (6.38,-1.1);
\draw[thick] (6.44,-1.5) -- (6.44,-1.1);
\draw[thick] (6.50,-1.5) -- (6.50,-1.1);
\draw[blue,very thick] (6.51,-1.3) -- (7.0,-1.3);
\node at (6.41,-0.95) {\scriptsize{FBG}};
\end{scope}
\end{scope}
\begin{scope}[shift={(3.5,-3.25)}]
\draw[thick] (4.4,0.2) rectangle (4.9,0.6);
\node at (4.65,0.73) {\scriptsize{BPF}};
\draw[thick] (4.5,0.25) .. controls (4.65,0.65) .. (4.8,0.25);
\end{scope}
\draw[blue,very thick] (7.2,-2.85) -- (7.9,-2.85);
\begin{scope}[shift={(0.1,-1.75)}]
\draw[thick] (4.3,-1.3) rectangle (4.9,-0.9);
\node at (4.60,-1.10) {\scriptsize{ESA}};
\end{scope}

\begin{scope}[rotate around = {90:(7.51,0.75)},shift={(-3.70,2.05)}]
\draw[thick,fill,black!90] (7.51,0.75) .. controls (7.56,0.95) and (7.66,0.95) .. (7.71,0.75);
\end{scope}
\draw[very thick,brown] (5.0,-2.85) -- (5.3,-2.85);

\draw[blue!40,fill,opacity=0.2,rounded corners] (4.34,-3.55) rectangle (7.25,-2.15);
\node at (5.8,-2.35) {\scriptsize{Phase detection}};

\draw[red!40,fill,opacity=0.2,rounded corners] (7.8,-3.5) -- (7.8,-2.3) -- (8.5,-2.3) -- (8.5,-3.1) -- (9.65,-3.1) -- (9.65,-3.5) -- cycle;
\node at (8.73,-3.3) {\scriptsize{Remove pump}};

\end{tikzpicture}
\endpgfgraphicnamed
}
\caption{\textbf{Experimental characterization of the photon-phonon coupling.} \textbf{a}, A typical Lorentzian gain profile on Stokes photons and (inset) a depletion profile on anti-Stokes photons. \textbf{b}, The fiber-based set-up used to monitor the forward Stokes scattering. A tunable laser is amplified in one arm by an erbium-doped fiber amplifier (EDFA) to serve as a pump. In the other arm, the laser light is intensity modulated (IM) to generate a Stokes and anti-Stokes sideband. Next, a fiber Bragg grating (FBG) rejects all but the Stokes line (see Methods). The pump and Stokes are coupled to the chip through curved grating couplers. Finally, the power in the Stokes and pump is monitored separately. With minor modifications, this set-up can be reconfigured to observe the anti-Stokes loss or backward scattering. The latter is weak in our wires (see supplementary information C), so here we focus on the forward scattering. \textbf{c}, A typical Fano signature obtained from the XPM-experiment. \textbf{d}, A pump is intensity modulated, amplified, combined with a probe wave and sent to the chip. At the output, the pump is removed. The phase modulation on the probe wave is transducted to intensity modulation by filtering out the anti-Stokes sideband. Finally, we use an electrical spectrum analyzer (ESA) to observe the imprinted tone.}
\vspace{-2mm}
\end{figure*}

First, we monitor the power in a Stokes wave as a function of frequency spacing with a strong pump wave (fig.2a-b). We observe a Lorentzian gain profile at $\frac{\Omega_{\text{m}}}{2\pi} = 9.2 \, \text{GHz}$, as expected in the low-cascading regime (see supplementary information A).  The Stokes photons experience exponential amplification as long as the pump remains undepleted. Exactly on resonance, the on/off gain is given by $2\gamma_{\text{SBS}}P_{\text{p}}L_{\text{eff}}$ -- with $\gamma_{\text{SBS}}$ the Brillouin nonlinearity parameter, $P_{\text{p}}$ the input pump power and $L_{\text{eff}}=\frac{1 - \exp{\left(- \alpha L \right)}}{\alpha}$ the effective interaction length. The effective length is limited to $\frac{1}{\alpha} = 1.7 \, \text{cm}$ in our wires. To extract the Brillouin nonlinearity $\gamma_{\text{SBS}}$, we sweep the pump power (fig.3a). Above powers of about $25 \, \text{mW}$, nonlinear absorption saturates the on/off gain. Then free carriers, created by two-photon absorption (TPA), result in a power-dependent optical loss $\alpha(P_{\text{p}})$. However, we extract $2\gamma_{\text{SBS}} = 3218 \, \text{W}^{-1}\text{m}^{-1}$ below this threshold. A fit to the Lorentzian resonance yields a mechanical quality factor of $Q_{\text{m}} = \frac{\Omega_{\text{m}}}{\Gamma_{\text{m}}} = 306$ for the same $2.7 \, \text{mm}$-long waveguide. Therefore the phonon lifetime is $\tau = \frac{1}{\Gamma_{\text{m}}} = 5.3 \, \text{ns}$. Besides, we note that the largest on/off gain of $0.6 \, \text{dB}$ below the TPA-threshold falls narrowly short of the linear loss $\alpha L = 0.7 \, \text{dB}$. Thus the wire is close to net optical amplification, which is necessary to make a Brillouin laser. The peak gain reaches $4.4 \, \text{dB}$ in the longest $4 \, \text{cm}$-wires (fig.2a), improving by a factor 11 on previous results in silicon\cite{Shin2013b}. Finally, we observe an identical depletion profile on an anti-Stokes wave (fig.2a).

Second, we measure the strength of the cross-phase modulation (XPM) imprinted on a weak probe by a strong intensity-modulated pump (fig.2c-d). The experiment yields a distinct asymmetric Fano signature at $\frac{\Omega_{\text{m}}}{2\pi} = 9.2 \; \text{GHz}$ caused by interference between the resonant Brillouin and the non-resonant Kerr response (see supplementary information B). The lineshape follows $|\frac{\gamma_{\text{XPM}}(\Omega)}{2\gamma_{\text{K}}}|^{2}$, with $\gamma_{\text{K}}$ the Kerr nonlinearity parameter and
\begin{gather*}
\gamma_{\text{XPM}}(\Omega) = 2\gamma_{\text{K}}+\gamma_{\text{SBS}}\mathcal{L}(\Omega) \notag \\
\mathcal{L}(\Omega) =\frac{1}{-2\Delta_r + i} \qquad \Delta_r = \frac{\Omega - \Omega_{\text{m}}}{\Gamma_{\text{m}}} \notag
\end{gather*}
From this resonance we isolate the ratio $\gamma_{\text{SBS}}/\gamma_{\text{K}} = 2.5$ and $Q_{\text{m}} = 249$. The Kerr parameter $\gamma_{\text{K}}$ of similar silicon wires has been studied extensively, with values reported at $\gamma_{\text{K}} = 566  \, \text{W}^{-1}\text{m}^{-1}$ for our cross-section\cite{Osgood2009}. Because of the pillar etch, the light is more confined to the high-index silicon core. We simulate that this results in a slight increase of the Kerr effect by $8\%$ to $\gamma_{\text{K}} = 611  \, \text{W}^{-1}\text{m}^{-1}$. Thus we have $2\gamma_{\text{SBS}} = 3055  \, \text{W}^{-1}\text{m}^{-1}$, within $5\%$ of the value obtained from the gain experiments.

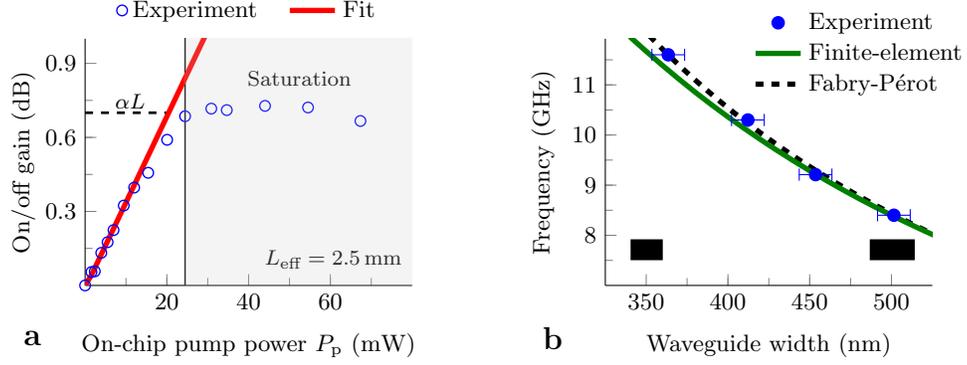
\begin{figure*}
\centering
\label{fig:2}
\subfigure{
%
%
%
%
\begin{tikzpicture}

\begin{axis}[scale = 0.28*\textwidth/(4.5in),
width=4.52083333333333in,
height=3.565625in,
xmin=0,
xmax=80,
xlabel={On-chip pump power $P_{\text{p}}$ (mW)},
ymin=0,
ymax=1,
ylabel={On/off gain (dB)},
ylabel style = {yshift=-3ex},
axis x line*=bottom,
axis y line*=left,
legend style={fill=none,draw=none,legend cell align=left,at={(0.5,1.2)},anchor=north, legend columns=-1},
xtick = {20,40,60},
ytick = {0.3,0.6,0.9},
minor xtick={10,20,...,70},
minor ytick={0.15,0.3,...,0.9}
]
\addplot [
color=black,
dashed,
line width=1.0pt,
forget plot
]
table[row sep=crcr]{
0 0.7\\
20.45 0.7\\
};
\addplot [
color=black,
solid,
line width=0.5pt,
forget plot
]
table[row sep=crcr]{
24.4620893987076 0\\
24.4620893987076 1.01 \\
};
\addplot [
color=blue,
line width=0.5pt,
mark size=2.0pt,
only marks,
mark=o,
mark options={solid},
]
table[row sep=crcr]{
0 0 \\
1.55444715242624 0.054287753331167\\
2.37257723265057 0.0571881068952938\\
3.95429538775095 0.132010806090318\\
5.50874254017718 0.175318390536979\\
6.95410568190684 0.224242013867835\\
9.46303792792813 0.323735349161231\\
11.9992411766236 0.396593055999814\\
15.4353875135658 0.456586786855763\\
20.0714579681703 0.590398829729755\\
24.4620893987076 0.685830654808906\\
30.8707750271315 0.716695957314655\\
34.6614443988376 0.710735559181219\\
44.0153983160692 0.72782369769756\\
54.542005348289 0.721318443345136\\
67.3593766051368 0.666672528068749\\
NaN 0.014581789801439\\
};
\addlegendentry{Experiment \, \,};

\addplot [
color=red,
solid,
line width=2.0pt,
]
table[row sep=crcr]{
0 0 \\
1.55444715242624 0.0422139914876601\\
1.65444715242624 0.0456977817730367\\
1.75444715242624 0.0491815720584132\\
1.85444715242624 0.0526653623437898\\
1.95444715242624 0.0561491526291664\\
2.05444715242623 0.059632942914543\\
2.15444715242624 0.0631167331999196\\
2.25444715242624 0.0666005234852962\\
2.35444715242624 0.0700843137706728\\
2.45444715242624 0.0735681040560494\\
2.55444715242624 0.077051894341426\\
2.65444715242624 0.0805356846268026\\
2.75444715242624 0.0840194749121791\\
2.85444715242624 0.0875032651975557\\
2.95444715242624 0.0909870554829323\\
3.05444715242623 0.0944708457683089\\
3.15444715242624 0.0979546360536855\\
3.25444715242624 0.101438426339062\\
3.35444715242624 0.104922216624439\\
3.45444715242624 0.108406006909815\\
3.55444715242624 0.111889797195192\\
3.65444715242624 0.115373587480568\\
3.75444715242624 0.118857377765945\\
3.85444715242624 0.122341168051322\\
3.95444715242624 0.125824958336698\\
4.05444715242623 0.129308748622075\\
4.15444715242624 0.132792538907451\\
4.25444715242624 0.136276329192828\\
4.35444715242623 0.139760119478205\\
4.45444715242624 0.143243909763581\\
4.55444715242624 0.146727700048958\\
4.65444715242624 0.150211490334334\\
4.75444715242624 0.153695280619711\\
4.85444715242624 0.157179070905088\\
4.95444715242624 0.160662861190464\\
5.05444715242623 0.164146651475841\\
5.15444715242624 0.167630441761217\\
5.25444715242624 0.171114232046594\\
5.35444715242623 0.17459802233197\\
5.45444715242624 0.178081812617347\\
5.55444715242624 0.181565602902724\\
5.65444715242624 0.1850493931881\\
5.75444715242624 0.188533183473477\\
5.85444715242624 0.192016973758853\\
5.95444715242624 0.19550076404423\\
6.05444715242624 0.198984554329607\\
6.15444715242624 0.202468344614983\\
6.25444715242624 0.20595213490036\\
6.35444715242624 0.209435925185736\\
6.45444715242624 0.212919715471113\\
6.55444715242624 0.21640350575649\\
6.65444715242624 0.219887296041866\\
6.75444715242624 0.223371086327243\\
6.85444715242624 0.226854876612619\\
6.95444715242624 0.230338666897996\\
7.05444715242624 0.233822457183373\\
7.15444715242624 0.237306247468749\\
7.25444715242624 0.240790037754126\\
7.35444715242624 0.244273828039502\\
7.45444715242624 0.247757618324879\\
7.55444715242624 0.251241408610256\\
7.65444715242624 0.254725198895632\\
7.75444715242624 0.258208989181009\\
7.85444715242624 0.261692779466385\\
7.95444715242623 0.265176569751762\\
8.05444715242624 0.268660360037138\\
8.15444715242623 0.272144150322515\\
8.25444715242623 0.275627940607892\\
8.35444715242624 0.279111730893268\\
8.45444715242624 0.282595521178645\\
8.55444715242623 0.286079311464021\\
8.65444715242624 0.289563101749398\\
8.75444715242624 0.293046892034775\\
8.85444715242623 0.296530682320151\\
8.95444715242624 0.300014472605528\\
9.05444715242623 0.303498262890904\\
9.15444715242623 0.306982053176281\\
9.25444715242624 0.310465843461658\\
9.35444715242624 0.313949633747034\\
9.45444715242624 0.317433424032411\\
9.55444715242623 0.320917214317787\\
9.65444715242623 0.324401004603164\\
9.75444715242624 0.327884794888541\\
9.85444715242624 0.331368585173917\\
9.95444715242624 0.334852375459294\\
10.0544471524262 0.33833616574467\\
10.1544471524262 0.341819956030047\\
10.2544471524262 0.345303746315423\\
10.3544471524262 0.3487875366008\\
10.4544471524262 0.352271326886177\\
10.5544471524262 0.355755117171553\\
10.6544471524262 0.35923890745693\\
10.7544471524262 0.362722697742306\\
10.8544471524262 0.366206488027683\\
10.9544471524262 0.36969027831306\\
11.0544471524262 0.373174068598436\\
11.1544471524262 0.376657858883813\\
11.2544471524262 0.380141649169189\\
11.3544471524262 0.383625439454566\\
11.4544471524262 0.387109229739943\\
11.5544471524262 0.390593020025319\\
11.6544471524262 0.394076810310696\\
11.7544471524262 0.397560600596072\\
11.8544471524262 0.401044390881449\\
11.9544471524262 0.404528181166825\\
12.0544471524262 0.408011971452202\\
12.1544471524262 0.411495761737579\\
12.2544471524262 0.414979552022955\\
12.3544471524262 0.418463342308332\\
12.4544471524262 0.421947132593708\\
12.5544471524262 0.425430922879085\\
12.6544471524262 0.428914713164462\\
12.7544471524262 0.432398503449838\\
12.8544471524262 0.435882293735215\\
12.9544471524262 0.439366084020591\\
13.0544471524262 0.442849874305968\\
13.1544471524262 0.446333664591345\\
13.2544471524262 0.449817454876721\\
13.3544471524262 0.453301245162098\\
13.4544471524262 0.456785035447474\\
13.5544471524262 0.460268825732851\\
13.6544471524262 0.463752616018228\\
13.7544471524262 0.467236406303604\\
13.8544471524262 0.470720196588981\\
13.9544471524262 0.474203986874357\\
14.0544471524262 0.477687777159734\\
14.1544471524262 0.48117156744511\\
14.2544471524262 0.484655357730487\\
14.3544471524262 0.488139148015864\\
14.4544471524262 0.49162293830124\\
14.5544471524262 0.495106728586617\\
14.6544471524262 0.498590518871993\\
14.7544471524262 0.50207430915737\\
14.8544471524262 0.505558099442747\\
14.9544471524262 0.509041889728123\\
15.0544471524262 0.5125256800135\\
15.1544471524262 0.516009470298876\\
15.2544471524262 0.519493260584253\\
15.3544471524262 0.52297705086963\\
15.4544471524262 0.526460841155006\\
15.5544471524262 0.529944631440383\\
15.6544471524262 0.533428421725759\\
15.7544471524262 0.536912212011136\\
15.8544471524262 0.540396002296513\\
15.9544471524262 0.543879792581889\\
16.0544471524262 0.547363582867266\\
16.1544471524262 0.550847373152642\\
16.2544471524262 0.554331163438019\\
16.3544471524262 0.557814953723396\\
16.4544471524262 0.561298744008772\\
16.5544471524262 0.564782534294149\\
16.6544471524262 0.568266324579525\\
16.7544471524262 0.571750114864902\\
16.8544471524262 0.575233905150278\\
16.9544471524262 0.578717695435655\\
17.0544471524262 0.582201485721032\\
17.1544471524262 0.585685276006408\\
17.2544471524262 0.589169066291785\\
17.3544471524262 0.592652856577161\\
17.4544471524262 0.596136646862538\\
17.5544471524262 0.599620437147915\\
17.6544471524262 0.603104227433291\\
17.7544471524262 0.606588017718668\\
17.8544471524262 0.610071808004044\\
17.9544471524262 0.613555598289421\\
18.0544471524262 0.617039388574797\\
18.1544471524262 0.620523178860174\\
18.2544471524262 0.624006969145551\\
18.3544471524262 0.627490759430927\\
18.4544471524262 0.630974549716304\\
18.5544471524262 0.63445834000168\\
18.6544471524262 0.637942130287057\\
18.7544471524262 0.641425920572434\\
18.8544471524262 0.64490971085781\\
18.9544471524262 0.648393501143187\\
19.0544471524262 0.651877291428563\\
19.1544471524262 0.65536108171394\\
19.2544471524262 0.658844871999317\\
19.3544471524262 0.662328662284693\\
19.4544471524262 0.66581245257007\\
19.5544471524262 0.669296242855446\\
19.6544471524262 0.672780033140823\\
19.7544471524262 0.6762638234262\\
19.8544471524262 0.679747613711576\\
19.9544471524262 0.683231403996953\\
20.0544471524262 0.686715194282329\\
20.1544471524262 0.690198984567706\\
20.2544471524262 0.693682774853083\\
20.3544471524262 0.697166565138459\\
20.4544471524262 0.700650355423836\\
20.5544471524262 0.704134145709212\\
20.6544471524262 0.707617935994589\\
20.7544471524262 0.711101726279965\\
20.8544471524262 0.714585516565342\\
20.9544471524262 0.718069306850719\\
21.0544471524262 0.721553097136095\\
21.1544471524262 0.725036887421472\\
21.2544471524262 0.728520677706848\\
21.3544471524262 0.732004467992225\\
21.4544471524262 0.735488258277601\\
21.5544471524262 0.738972048562978\\
21.6544471524262 0.742455838848355\\
21.7544471524262 0.745939629133731\\
21.8544471524262 0.749423419419108\\
21.9544471524262 0.752907209704484\\
22.0544471524262 0.756390999989861\\
22.1544471524262 0.759874790275238\\
22.2544471524262 0.763358580560614\\
22.3544471524262 0.766842370845991\\
22.4544471524262 0.770326161131367\\
22.5544471524262 0.773809951416744\\
22.6544471524262 0.777293741702121\\
22.7544471524262 0.780777531987497\\
22.8544471524262 0.784261322272874\\
22.9544471524262 0.78774511255825\\
23.0544471524262 0.791228902843627\\
23.1544471524262 0.794712693129004\\
23.2544471524262 0.79819648341438\\
23.3544471524262 0.801680273699757\\
23.4544471524262 0.805164063985133\\
23.5544471524262 0.80864785427051\\
23.6544471524262 0.812131644555886\\
23.7544471524262 0.815615434841263\\
23.8544471524262 0.81909922512664\\
23.9544471524262 0.822583015412016\\
24.0544471524262 0.826066805697393\\
24.1544471524262 0.829550595982769\\
24.2544471524262 0.833034386268146\\
24.3544471524262 0.836518176553523\\
24.4544471524262 0.840001966838899\\
24.5544471524262 0.843485757124276\\
24.6544471524262 0.846969547409652\\
24.7544471524262 0.850453337695029\\
24.8544471524262 0.853937127980406\\
24.9544471524262 0.857420918265782\\
25.0544471524262 0.860904708551159\\
25.1544471524262 0.864388498836535\\
25.2544471524262 0.867872289121912\\
25.3544471524262 0.871356079407289\\
25.4544471524262 0.874839869692665\\
25.5544471524262 0.878323659978042\\
25.6544471524262 0.881807450263418\\
25.7544471524262 0.885291240548795\\
25.8544471524262 0.888775030834172\\
25.9544471524262 0.892258821119548\\
26.0544471524262 0.895742611404925\\
26.1544471524262 0.899226401690301\\
26.2544471524262 0.902710191975678\\
26.3544471524262 0.906193982261054\\
26.4544471524262 0.909677772546431\\
26.5544471524262 0.913161562831808\\
26.6544471524262 0.916645353117184\\
26.7544471524262 0.920129143402561\\
26.8544471524262 0.923612933687937\\
26.9544471524262 0.927096723973314\\
27.0544471524262 0.930580514258691\\
27.1544471524262 0.934064304544067\\
27.2544471524262 0.937548094829444\\
27.3544471524262 0.94103188511482\\
27.4544471524262 0.944515675400197\\
27.5544471524262 0.947999465685573\\
27.6544471524262 0.95148325597095\\
27.7544471524262 0.954967046256327\\
27.8544471524262 0.958450836541703\\
27.9544471524262 0.96193462682708\\
28.0544471524262 0.965418417112456\\
28.1544471524262 0.968902207397833\\
28.2544471524262 0.97238599768321\\
28.3544471524262 0.975869787968586\\
28.4544471524262 0.979353578253963\\
28.5544471524262 0.982837368539339\\
28.6544471524262 0.986321158824716\\
28.7544471524262 0.989804949110093\\
28.8544471524262 0.993288739395469\\
28.9544471524262 0.996772529680846\\
29.0544471524262 1.00025631996622\\
29.1544471524262 1.0037401102516\\
29.2544471524262 1.00722390053698\\
29.3544471524262 1.01070769082235\\
29.4544471524262 1.01419148110773\\
29.5544471524262 1.01767527139311\\
29.6544471524262 1.02115906167848\\
29.7544471524262 1.02464285196386\\
29.8544471524262 1.02812664224924\\
29.9544471524262 1.03161043253461\\
30.0544471524262 1.03509422281999\\
30.1544471524262 1.03857801310537\\
30.2544471524262 1.04206180339074\\
30.3544471524262 1.04554559367612\\
30.4544471524262 1.04902938396149\\
30.5544471524262 1.05251317424687\\
30.6544471524262 1.05599696453225\\
30.7544471524262 1.05948075481762\\
30.8544471524262 1.062964545103\\
30.9544471524262 1.06644833538838\\
31.0544471524262 1.06993212567375\\
31.1544471524262 1.07341591595913\\
31.2544471524262 1.07689970624451\\
31.3544471524262 1.08038349652988\\
31.4544471524262 1.08386728681526\\
31.5544471524262 1.08735107710064\\
31.6544471524262 1.09083486738601\\
31.7544471524262 1.09431865767139\\
31.8544471524262 1.09780244795677\\
31.9544471524262 1.10128623824214\\
32.0544471524262 1.10477002852752\\
32.1544471524262 1.1082538188129\\
32.2544471524262 1.11173760909827\\
32.3544471524262 1.11522139938365\\
32.4544471524262 1.11870518966903\\
32.5544471524262 1.1221889799544\\
32.6544471524262 1.12567277023978\\
32.7544471524262 1.12915656052516\\
32.8544471524262 1.13264035081053\\
32.9544471524262 1.13612414109591\\
33.0544471524262 1.13960793138129\\
33.1544471524262 1.14309172166666\\
33.2544471524262 1.14657551195204\\
33.3544471524262 1.15005930223742\\
33.4544471524262 1.15354309252279\\
33.5544471524262 1.15702688280817\\
33.6544471524262 1.16051067309355\\
33.7544471524262 1.16399446337892\\
33.8544471524262 1.1674782536643\\
33.9544471524262 1.17096204394968\\
34.0544471524262 1.17444583423505\\
34.1544471524262 1.17792962452043\\
34.2544471524262 1.18141341480581\\
34.3544471524262 1.18489720509118\\
34.4544471524262 1.18838099537656\\
34.5544471524262 1.19186478566193\\
34.6544471524262 1.19534857594731\\
34.7544471524262 1.19883236623269\\
34.8544471524262 1.20231615651806\\
34.9544471524262 1.20579994680344\\
35.0544471524262 1.20928373708882\\
35.1544471524262 1.21276752737419\\
35.2544471524262 1.21625131765957\\
35.3544471524262 1.21973510794495\\
35.4544471524262 1.22321889823032\\
35.5544471524262 1.2267026885157\\
35.6544471524262 1.23018647880108\\
35.7544471524262 1.23367026908645\\
35.8544471524262 1.23715405937183\\
35.9544471524262 1.24063784965721\\
36.0544471524262 1.24412163994258\\
36.1544471524262 1.24760543022796\\
36.2544471524262 1.25108922051334\\
36.3544471524262 1.25457301079871\\
36.4544471524262 1.25805680108409\\
36.5544471524262 1.26154059136947\\
36.6544471524262 1.26502438165484\\
36.7544471524262 1.26850817194022\\
36.8544471524262 1.2719919622256\\
36.9544471524262 1.27547575251097\\
37.0544471524262 1.27895954279635\\
37.1544471524262 1.28244333308173\\
37.2544471524262 1.2859271233671\\
37.3544471524262 1.28941091365248\\
37.4544471524262 1.29289470393786\\
37.5544471524262 1.29637849422323\\
37.6544471524262 1.29986228450861\\
37.7544471524262 1.30334607479399\\
37.8544471524262 1.30682986507936\\
37.9544471524262 1.31031365536474\\
38.0544471524262 1.31379744565012\\
38.1544471524262 1.31728123593549\\
38.2544471524262 1.32076502622087\\
38.3544471524262 1.32424881650625\\
38.4544471524262 1.32773260679162\\
38.5544471524262 1.331216397077\\
38.6544471524262 1.33470018736238\\
38.7544471524262 1.33818397764775\\
38.8544471524262 1.34166776793313\\
38.9544471524262 1.34515155821851\\
39.0544471524262 1.34863534850388\\
39.1544471524262 1.35211913878926\\
39.2544471524262 1.35560292907463\\
39.3544471524262 1.35908671936001\\
39.4544471524262 1.36257050964539\\
39.5544471524262 1.36605429993076\\
39.6544471524262 1.36953809021614\\
39.7544471524262 1.37302188050152\\
39.8544471524262 1.37650567078689\\
39.9544471524262 1.37998946107227\\
};
\addlegendentry{Fit};
\end{axis}
\node[draw=none] at (3.3,0.35) {\footnotesize $L_{\text{eff}}=2.5 \,\text{mm}$};
\node[draw=none] at (2.85,2.75) {\footnotesize Saturation};
\node[draw=none] at (0.6,2.45) {\footnotesize $\alpha L$};
\node at (-0.7,-0.7) {\large{\textbf{a}}};
\draw[black!20,fill,opacity=0.2] (1.32,0) rectangle (4.34,3.29);
\end{tikzpicture}%
}
\subfigure{
\hspace{10mm}
%
%
%
%
\begin{tikzpicture}

\begin{axis}[scale = 0.28*\textwidth/(4.5in),
width=4.52083333333333in,
height=3.565625in,
xmin=325,
xmax=525,
xlabel={Waveguide width (nm)},
ymin=7,
ymax=11.92,
ylabel={Frequency (GHz)},
ylabel style = {yshift=-3ex},
axis x line*=bottom,
axis y line*=left,
reverse legend,
legend style={fill=none,draw=none,legend cell align=left,at={(0.78,1.15)},anchor=north},
xtick = {350,400,450,500},
ytick = {8,9,10,11},
minor xtick={375,400,...,500},
minor ytick={7.5,8,...,11.5}
]
\addplot [
color=black,
dashed,
line width = 2.0pt
]
table[row sep=crcr]{
320 13.1765625\\
321 13.1355140186916\\
322 13.0947204968944\\
323 13.0541795665635\\
324 13.0138888888889\\
325 12.9738461538462\\
326 12.9340490797546\\
327 12.894495412844\\
328 12.8551829268293\\
329 12.8161094224924\\
330 12.7772727272727\\
331 12.738670694864\\
332 12.7003012048193\\
333 12.6621621621622\\
334 12.624251497006\\
335 12.5865671641791\\
336 12.5491071428571\\
337 12.5118694362018\\
338 12.4748520710059\\
339 12.4380530973451\\
340 12.4014705882353\\
341 12.3651026392962\\
342 12.3289473684211\\
343 12.2930029154519\\
344 12.2572674418605\\
345 12.2217391304348\\
346 12.1864161849711\\
347 12.1512968299712\\
348 12.1163793103448\\
349 12.0816618911175\\
350 12.0471428571429\\
351 12.0128205128205\\
352 11.9786931818182\\
353 11.9447592067989\\
354 11.9110169491525\\
355 11.8774647887324\\
356 11.8441011235955\\
357 11.8109243697479\\
358 11.7779329608939\\
359 11.7451253481894\\
360 11.7125\\
361 11.6800554016621\\
362 11.6477900552486\\
363 11.6157024793388\\
364 11.5837912087912\\
365 11.5520547945205\\
366 11.5204918032787\\
367 11.4891008174387\\
368 11.4578804347826\\
369 11.4268292682927\\
370 11.3959459459459\\
371 11.3652291105121\\
372 11.3346774193548\\
373 11.3042895442359\\
374 11.274064171123\\
375 11.244\\
376 11.2140957446809\\
377 11.184350132626\\
378 11.1547619047619\\
379 11.1253298153034\\
380 11.0960526315789\\
381 11.0669291338583\\
382 11.0379581151832\\
383 11.009138381201\\
384 10.98046875\\
385 10.9519480519481\\
386 10.9235751295337\\
387 10.8953488372093\\
388 10.8672680412371\\
389 10.8393316195373\\
390 10.8115384615385\\
391 10.7838874680307\\
392 10.7563775510204\\
393 10.7290076335878\\
394 10.7017766497462\\
395 10.6746835443038\\
396 10.6477272727273\\
397 10.6209068010076\\
398 10.5942211055276\\
399 10.5676691729323\\
400 10.54125\\
401 10.5149625935162\\
402 10.4888059701493\\
403 10.4627791563275\\
404 10.4368811881188\\
405 10.4111111111111\\
406 10.3854679802956\\
407 10.3599508599509\\
408 10.3345588235294\\
409 10.3092909535452\\
410 10.2841463414634\\
411 10.2591240875912\\
412 10.2342233009709\\
413 10.2094430992736\\
414 10.1847826086957\\
415 10.1602409638554\\
416 10.1358173076923\\
417 10.1115107913669\\
418 10.0873205741627\\
419 10.063245823389\\
420 10.0392857142857\\
421 10.0154394299287\\
422 9.99170616113744\\
423 9.96808510638298\\
424 9.94457547169811\\
425 9.92117647058824\\
426 9.89788732394366\\
427 9.87470725995316\\
428 9.85163551401869\\
429 9.82867132867133\\
430 9.80581395348837\\
431 9.7830626450116\\
432 9.76041666666667\\
433 9.7378752886836\\
434 9.71543778801843\\
435 9.69310344827586\\
436 9.67087155963303\\
437 9.6487414187643\\
438 9.62671232876712\\
439 9.60478359908884\\
440 9.58295454545454\\
441 9.56122448979592\\
442 9.539592760181\\
443 9.51805869074492\\
444 9.49662162162162\\
445 9.4752808988764\\
446 9.45403587443946\\
447 9.43288590604027\\
448 9.41183035714286\\
449 9.39086859688196\\
450 9.37\\
451 9.34922394678492\\
452 9.32853982300885\\
453 9.30794701986755\\
454 9.28744493392071\\
455 9.26703296703297\\
456 9.24671052631579\\
457 9.22647702407002\\
458 9.20633187772926\\
459 9.18627450980392\\
460 9.16630434782609\\
461 9.14642082429501\\
462 9.12662337662338\\
463 9.10691144708423\\
464 9.08728448275862\\
465 9.06774193548387\\
466 9.04828326180258\\
467 9.02890792291221\\
468 9.00961538461539\\
469 8.99040511727079\\
470 8.97127659574468\\
471 8.95222929936306\\
472 8.93326271186441\\
473 8.91437632135307\\
474 8.89556962025316\\
475 8.87684210526316\\
476 8.85819327731092\\
477 8.83962264150943\\
478 8.82112970711297\\
479 8.8027139874739\\
480 8.784375\\
481 8.76611226611227\\
482 8.74792531120332\\
483 8.72981366459627\\
484 8.71177685950413\\
485 8.69381443298969\\
486 8.67592592592593\\
487 8.65811088295688\\
488 8.64036885245902\\
489 8.62269938650307\\
490 8.60510204081633\\
491 8.58757637474542\\
492 8.57012195121951\\
493 8.552738336714\\
494 8.53542510121457\\
495 8.51818181818182\\
496 8.50100806451613\\
497 8.48390342052314\\
498 8.46686746987952\\
499 8.4498997995992\\
500 8.433\\
501 8.41616766467066\\
502 8.39940239043825\\
503 8.38270377733598\\
504 8.36607142857143\\
505 8.34950495049505\\
506 8.33300395256917\\
507 8.31656804733728\\
508 8.3001968503937\\
509 8.28388998035364\\
510 8.26764705882353\\
511 8.25146771037182\\
512 8.2353515625\\
513 8.21929824561403\\
514 8.20330739299611\\
515 8.1873786407767\\
516 8.17151162790698\\
517 8.15570599613153\\
518 8.13996138996139\\
519 8.1242774566474\\
520 8.10865384615385\\
521 8.09309021113244\\
522 8.07758620689655\\
523 8.06214149139579\\
524 8.04675572519084\\
525 8.03142857142857\\
526 8.01615969581749\\
527 8.00094876660341\\
528 7.98579545454545\\
529 7.97069943289225\\
530 7.95566037735849\\
531 7.94067796610169\\
532 7.92575187969925\\
533 7.9108818011257\\
534 7.89606741573034\\
535 7.88130841121495\\
536 7.86660447761194\\
537 7.85195530726257\\
538 7.83736059479554\\
539 7.82282003710575\\
540 7.80833333333333\\
541 7.79390018484288\\
542 7.77952029520295\\
543 7.76519337016575\\
544 7.75091911764706\\
545 7.73669724770642\\
546 7.72252747252747\\
547 7.70840950639854\\
548 7.69434306569343\\
549 7.68032786885246\\
550 7.66636363636364\\
};
\addlegendentry{Fabry-P\'{e}rot};

\addplot [
color=green!50!black,
solid,
line width = 2.0pt,
mark options={solid},
]
table[row sep=crcr]{
340 11.959997089081\\
350 11.6698950442607\\
360 11.3880625556639\\
370 11.1156909277059\\
380 10.8535839462149\\
390 10.6026776463083\\
400 10.3619109504945\\
410 10.1304372844193\\
420 9.90779267437816\\
430 9.69347324445291\\
440 9.48715400888613\\
450 9.28840562578722\\
460 9.09686529817879\\
470 8.91251725455307\\
480 8.7352189628294\\
490 8.56468488712868\\
500 8.40049913766562\\
510 8.24226611147115\\
520 8.08966707694027\\
530 7.94243037349691\\
540 7.80030138363508\\
};
\addlegendentry{Finite-element};

\addplot [
color=blue,
line width=0.5pt,
mark size=2.3pt,
only marks,
mark=*,
mark options={solid},
]
plot[/pgfplots/error bars/.cd,
x dir=both,
y dir=none,
x explicit,
y explicit,
error mark=-,
error bar style={color=blue}]
table[x=x,y=y,x error=errx,y error=erry,row sep=crcr]{
x y errx erry \\
363.5 11.6 10 0.1\\
412.2 10.3 10 0.1\\
453.6 9.21 10 0.1\\
501.5 8.4 10 0.1 \\
};
\addlegendentry{Experiment};
\end{axis}
\node at (-0.7,-0.7) {\large{\textbf{b}}};
\draw[black,thick,fill] (0.35,0.35) rectangle (0.75,0.6) ;
\begin{scope}[shift={(3.18,0.0)}]
\draw[black,thick,fill] (0.35,0.35) rectangle (0.92,0.6) ;
\end{scope}
\end{tikzpicture}%
}
\caption{\textbf{Analysis of the Brillouin gain and phonon frequency.} \textbf{a}, Scaling of the on/off Brillouin gain with input pump power. Above a power threshold of $25 \, \text{mW}$, the on/off gain saturates because of nonlinear absorption. We perform a fit to obtain the Brillouin nonlinearity below that threshold. \textbf{b}, The phonon frequency for different waveguide widths. Both a simple Fabry-P\'{e}rot and a rigorous finite-element model agree with the data.}
\end{figure*}

We stress that both the resonance frequency, quality factor and coupling strength are in excellent agreement with the models. For the frequency, we perform the XPM-experiment for waveguide widths from $350 \, \text{nm}$ to $500 \, \text{nm}$ (fig.3b). Both a simple Fabry-P\'{e}rot ($\frac{\Omega_{\text{m}}}{2\pi} = \frac{v}{2w}$) and a sophisticated finite-element model match the observed resonances. The finite-element model takes into account the exact geometry of the wires as obtained from a scanning electron micrograph (fig.1c). This includes the waveguide height, pillar size, sidewall angle and the $\langle 110 \rangle$ crystal orientation of our wires (see Methods). We find that the waveguide width alone pins down the resonance frequency, with other geometrical parameters inducing minor shifts. For a $450 \, \text{nm}$-wide waveguide, the frequency sensitivity to width changes is $19.2 \, \text{MHz/nm}$ (fig.3b). In contrast, the calculated sensitivity to height changes is only $2.3 \, \text{MHz/nm}$. This is consistent with the intuitive Fabry-P\'{e}rot picture, in which the height does not appear at all.

The large sensitivity to width variations implies that a $2 \, \text{nm}$ width change shifts the resonance by more than a linewidth. Therefore inhomogeneous broadening may affect both the lineshape and -width in the longer wires, similar to Doppler-broadening in gain media. Surprisingly, we achieve quality factors above $250$ even in the longest $4 \, \text{cm}$-wires (fig.4a). This suggests that there is, if at all, only limited length-dependent line broadening.

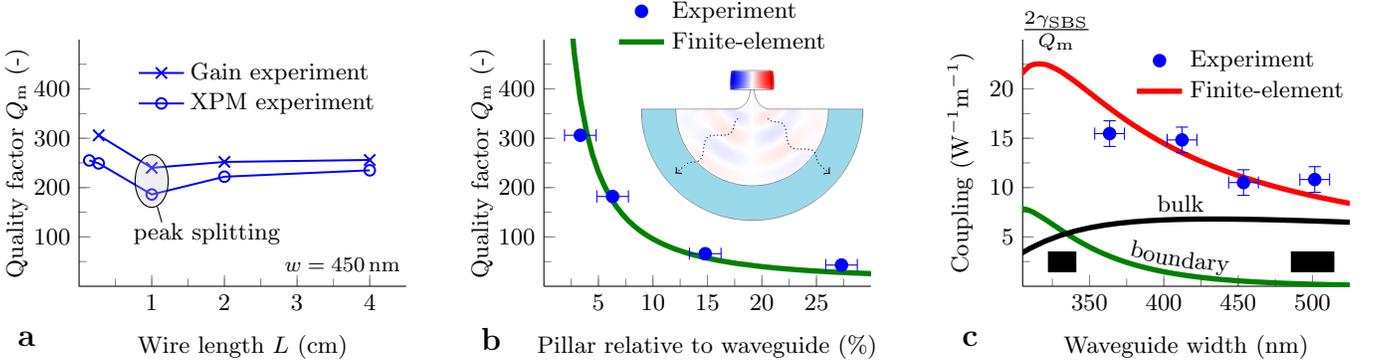
\begin{figure*}
\centering
\label{fig:2}
\subfigure{\hspace{-7mm}
\begin{tikzpicture}

\begin{axis}[scale = 0.28*\textwidth/(4.5in),
width=4.52083333333333in,
height=3.565625in,
xmin=0,
xmax=4.5,
xlabel={Wire length $L$ (cm)},
ymin=0,
ymax=500,
ylabel={Quality factor $Q_{\text{m}}$ (-)},
ylabel style = {yshift=-3ex},
axis x line*=bottom,
axis y line*=left,
legend style={fill=none,draw=none,legend cell align=left,at={(0.55,0.95)},anchor=north},
xtick = {1,2,...,4},
ytick = {100,200,300,400},
minor xtick={0.5,1,...,4},
minor ytick={50,100,...,450}
]
\addplot [
color=blue,
line width=0.75pt,
mark size=3.0pt,
solid,
mark=x,
mark options={solid},
]
table[row sep=crcr]{
0.27 306 \\
1 240 \\
2 252 \\
4 256 \\
};
\addlegendentry{Gain experiment};

\addplot [
color=blue,
line width=0.75pt,
mark size=2.0pt,
solid,
mark=o,
mark options={solid},
]
table[row sep=crcr]{
0.14 255 \\
0.27 249 \\
1 186 \\
2 222 \\
4 235 \\
};
\addlegendentry{XPM experiment};

\end{axis}
\node at (-0.7,-0.7) {\large{\textbf{a}}};
\draw[black!20,fill,opacity=0.3] (0.965,1.4) ellipse (2.2mm and 3.5mm);
\draw[black] (0.965,1.4) ellipse (2.2mm and 3.5mm);
\draw[black] (1.04,1.06) -- (1.1,0.85);
\node[draw=none] at (3.5,0.29) {\footnotesize $w =  450 \, \text{nm}$};
\node[draw=none] at (1.7,0.70) {peak splitting};
\end{tikzpicture}}
\subfigure{\hspace{5mm}
\begin{tikzpicture}

\begin{axis}[scale = 0.28*\textwidth/(4.5in),
width=4.52083333333333in,
height=3.565625in,
xmin=0,
xmax=30,
xlabel={Pillar relative to waveguide (\%)},
ymin=0,
ymax=500,
ylabel={Quality factor $Q_{\text{m}}$ (-)},
ylabel style = {yshift=-3ex},
axis x line*=bottom,
axis y line*=left,
reverse legend, 
legend style={fill=none,draw=none,legend cell align=left,at={(0.55,1.20)},anchor=north},
xtick = {5,10,...,20,25},
ytick = {100,200,300,400},
minor xtick={2.5,5,...,27.5},
minor ytick={50,100,...,450}
]
\addplot [
color=green!50!black,
line width = 2.0pt
]
table[row sep=crcr]{
2.21680336954112 615.368017278162\\
2.7710042119264 474.28881094207\\
3.32520505431168 380.777718689724\\
3.87940589669696 316.317485853393\\
4.43360673908224 268.184918448093\\
4.98780758146752 231.923393754894\\
5.5420084238528 203.741041228284\\
6.09620926623808 180.604856336506\\
6.65041010862337 161.493561152199\\
7.20461095100865 146.0474079463\\
7.75881179339393 132.577070309916\\
8.31301263577921 121.516304483229\\
8.86721347816449 111.824223905289\\
9.42141432054977 103.429797583519\\
9.97561516293505 96.0954593762624\\
10.5298160053203 89.6768116402945\\
11.0840168477056 83.9221653062311\\
11.6382176900909 78.8886836541947\\
12.1924185324762 74.1160557053197\\
12.7466193748615 70.0662354922297\\
13.3008202172467 66.2268664226636\\
13.855021059632 62.8621700619355\\
14.4092219020173 59.7428868758111\\
14.9634227444026 57.1139186378372\\
15.5176235867879 54.5950486429168\\
16.0718244291731 52.1199849712864\\
16.6260252715584 50.054762275635\\
17.1802261139437 48.014298700539\\
17.734426956329 46.0890651921031\\
18.2886277987143 44.5557097937848\\
18.8428286410995 42.9999003252996\\
19.3970294834848 41.4168340923192\\
19.9512303258701 39.9543146430818\\
20.5054311682554 38.7509348962983\\
21.0596320106407 37.4668413713141\\
21.6138328530259 36.3607737898239\\
22.1680336954112 35.2779044446221\\
22.7222345377965 34.3284004206098\\
23.2764353801818 33.3841260601999\\
23.8306362225671 32.4885204641487\\
24.3848370649523 31.6605936007383\\
24.9390379073376 30.9306797959962\\
25.4932387497229 30.1798553294897\\
26.0474395921082 29.4699850293433\\
26.6016404344935 28.8086611635556\\
27.1558412768787 28.2241090073941\\
27.710042119264 27.6685672873865\\
28.2642429616493 27.0836315011341\\
28.8184438040346 26.5927950085909\\
29.3726446464199 26.1348348129689\\
29.9268454888051 25.6646836093438\\
30.4810463311904 25.206006203877\\
31.0352471735757 24.8038059868832\\
31.589448015961 24.4028221749302\\
32.1436488583463 23.9477532022398\\
32.6978497007315 23.6073864078011\\
33.2520505431168 23.3582626898479\\
};
\addlegendentry{Finite-element};

\addplot [
color=blue,
line width=0.5pt,
mark size=2.3pt,
only marks,
mark=*,
mark options={solid}
]
plot[/pgfplots/error bars/.cd,
x dir=both,
y dir=none,
x explicit,
y explicit,
error mark=-,
error bar style={color=blue}]
table[x=x,y=y,x error=errx,y error = erry,row sep=crcr]{
x y errx erry\\
3.34 306 1.45 20\\
6.3 182 1.45 15\\
14.8 66 1.45 10\\
27.3 43 1.45 10\\
};
\addlegendentry{Experiment \, \,};
\end{axis}
\node at (-0.7,-0.7) {\large{\textbf{b}}};
\begin{scope}[shift={(+0.0,-0.15)}]
\node[anchor=south west,inner sep=0] at (1.2,1.0) {\raisebox{0pt}{\includegraphics[scale=0.175]{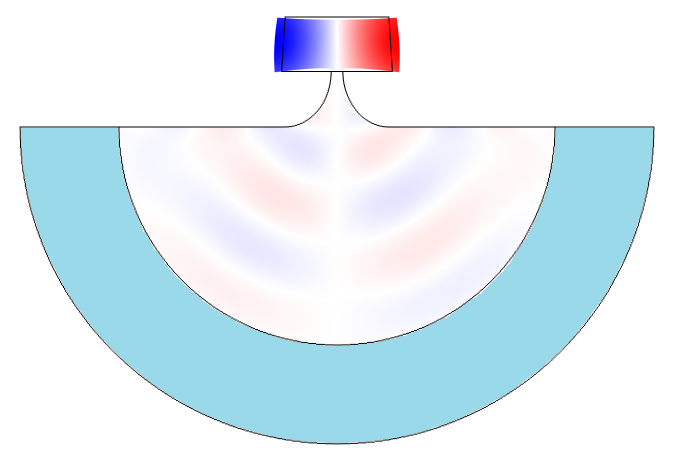}}};
\draw[->,thin,densely dotted,decorate,decoration={snake,amplitude=0.8mm,segment length=5.5mm,pre length=0.25mm,post length=0.5mm}] (2.58,2.38) -- (1.75,1.65);
\begin{scope}[shift={(+1.2,0.05)}]
\begin{scope}[rotate around={97:(2.2,2.0)}]
\draw[->,thin,densely dotted,decorate,decoration={snake,amplitude=0.8mm,segment length=5.5mm,pre length=0.25mm,post length=0.5mm}] (2.58,2.38) -- (1.75,1.65);
\end{scope}
\end{scope}
\end{scope}
\end{tikzpicture}}
\subfigure{\hspace{5mm}
\begin{tikzpicture}
\begin{axis}[scale = 0.28*\textwidth/(4.5in),
width=4.52083333333333in,
height=3.565625in,
xmin=305,
xmax=525,
xlabel={Waveguide width (nm)},
ymin=0,
ymax=25,
ylabel={Coupling ($\text{W}^{-1}\text{m}^{-1}$)},
ylabel style = {yshift=-3ex},
axis x line*=bottom,
axis y line*=left,
reverse legend,
legend style={fill=none,draw=none,legend cell align=left,at={(0.67,1.0)},anchor=north},
xtick = {350,400,450,500},
ytick = {5,10,15,20},
minor xtick={325,350,...,500},
minor ytick={2.5,5,...,22.5}
]
\addplot [
color=green!50!black,
smooth,
solid,
line width = 2.0pt,
mark options={solid},
]
table[row sep=crcr]{
300 8.01978580704474\\
310 7.65607276719619\\
320 6.79026370928843\\
330 5.78333953920683\\
340 4.82075238576202\\
350 3.97653510407816\\
360 3.26203767923144\\
370 2.67170467866642\\
380 2.18868519334832\\
390 1.79384794477328\\
400 1.46960979274498\\
410 1.20696408345647\\
420 0.993906970375256\\
430 0.815975551747494\\
440 0.671192799494569\\
450 0.55284783706567\\
460 0.455778269468719\\
470 0.375496903421166\\
480 0.308101420688968\\
490 0.25306222155349\\
500 0.207022167930653\\
510 0.169163948117127\\
520 0.137484947714538\\
530 0.111288050589617\\
540 0.0895634186144109\\
550 0.0715668220629418\\
};

\addplot [
color=black,
solid,
line width = 2.0pt,
mark options={solid},
]
table[row sep=crcr]{
300 2.98614223559277\\
310 3.84191466868697\\
320 4.56061161757351\\
330 5.13467271484798\\
340 5.58262774062089\\
350 5.92770766696345\\
360 6.19119158949798\\
370 6.39033490328517\\
380 6.53881430677952\\
390 6.64683909961174\\
400 6.72249419931957\\
410 6.77247300897987\\
420 6.80160099250299\\
430 6.81327792114246\\
440 6.81120595910135\\
450 6.7976862307833\\
460 6.77495667697057\\
470 6.74437786256695\\
480 6.70740470385378\\
490 6.66536483030052\\
500 6.61905542904563\\
510 6.56925588816023\\
520 6.51669375304199\\
530 6.46181417927657\\
540 6.40522119365286\\
550 6.34729149763951\\
};

\addplot [
color=red,
smooth,
solid,
line width = 2.0pt,
mark options={solid},
]
table[row sep=crcr]{
300 20.7933119870475\\
310 22.3449184799759\\
320 22.4806112732299\\
330 21.81673800208\\
340 20.7788253697292\\
350 19.6143895514756\\
360 18.441199060464\\
370 17.3259581825556\\
380 16.2935833102115\\
390 15.3467477617773\\
400 14.4784203401309\\
410 13.6975247720779\\
420 12.995568976428\\
430 12.3449583877095\\
440 11.7586739581701\\
450 11.2276908249273\\
460 10.7452067587919\\
470 10.3026364616833\\
480 9.89061371793344\\
490 9.51592741290912\\
500 9.1672663587588\\
510 8.84676635640432\\
520 8.54726855589667\\
530 8.26912428548226\\
540 8.00960936395972\\
550 7.76682787602513\\
};
\addlegendentry{Finite-element};

\addplot [
color=blue,
line width=0.5pt,
mark size=2.3pt,
only marks,
mark=*,
mark options={solid},
]
plot[/pgfplots/error bars/.cd,
x dir=both,
y dir=both,
x explicit,
y explicit,
error mark=-,
error bar style={color=blue}]
table[x=x,y=y,x error=errx,y error=erry,row sep=crcr]{
x y errx erry \\
363.5 15.47 10 1.3\\
412.2 14.83 10 1.3\\
453.6 10.51 10 1.3\\
501.5 10.82 10 1.3 \\
};
\addlegendentry{Experiment};
\end{axis}
\node[scale=1.2] at (0.45,3.4) {$\frac{2\gamma_{\text{SBS}}}{Q_{\text{m}}}$};
\node at (2.1,1.1) {{bulk}};

\path[draw=white,
      postaction={decorate},
      decoration={text along path,
                  text=boundary,
                  text align={left indent={0.01\dimexpr\pgfdecoratedpathlength\relax}}
                  }
                  ] 
(1.4,0.41) .. controls (2.0,0.26) .. (2.7,0.195);

\node at (-0.7,-0.7) {\large{\textbf{c}}};
\begin{scope}[shift={(0.0,-0.15)}]
\draw[black,thick,fill] (0.35,0.35) rectangle (0.7,0.6) ;
\begin{scope}[shift={(3.23,0.0)}]
\draw[black,thick,fill] (0.35,0.35) rectangle (0.90,0.6) ;
\end{scope}
\end{scope}
\end{tikzpicture}}
\caption{\textbf{Study of the mechanical quality factor and the intrinsic photon-phonon coupling.} \textbf{a}, The quality factor stays above $250$ even in $4 \, \text{cm}$-long spirals, showing no evidence of length-dependent line broadening. There is peak splitting (by $50 \, \text{MHz}$) in the $1 \, \text{cm}$ spiral. Neither the longer spirals nor the straight wires exhibit such splitting. \textbf{b}, A finite-element model of phonon leakage through the pillar accurately predicts the observed quality factors. \textbf{c}, The non-resonant nonlinearity $\frac{2\gamma_{\text{SBS}}}{Q_{\text{m}}}$ is a direct measure of the intrinsic photon-phonon overlap. The electrostriction (black) and radiation pressure (green) interfere constructively, bringing about the total overlap (red). As the width increases, the boundary contribution vanishes rapidly.}
\end{figure*}

By sweeping the pillar width in a short $450 \, \text{nm}$-wide waveguide, we establish leakage through the pillar as the dominant phononic loss mechanism (fig.4b). The pillar acts as a channel for elastic waves that propagate down into the substrate. We rigorously model this mechanism by adding an artificial absorbing layer at the boundary of the simulation domain (see Methods). As predicted by such a model, the observed quality factors diminish rapidly with increasing pillar size. The pillar should be seen as a moving acoustic membrane \cite{Anetsberger2008a}, not as a fixed point. Therefore it affects neither the phononic field profile nor its associated stiffness $k_{\text{eff}}$ considerably.

Finally, we confirm that the photon-phonon coupling is determined by a combination of bulk (electrostriction) and boundary (radiation pressure) effects (fig.4c). The resonant Brillouin gain coefficient is given by $G_{\text{SBS}}(\Omega_{\text{m}}) = 2\gamma_{\text{SBS}}=\omega_{\text{0}} Q_{\text{m}}|\langle \mathbf{f}, \mathbf{u}\rangle|^{2}/(2k_{\text{eff}})$, so the non-resonant part $\frac{2\gamma_{\text{SBS}}}{Q_{\text{m}}}$ is proportional to the intrinsic photon-phonon coupling \cite{Rakich2012,Qiu2013}. In our finite-element simulations of $\langle \mathbf{f}, \mathbf{u}\rangle$ and $k_{\text{eff}}$, we take into account the mechanical anisotropy of silicon but not the pillar. We also approximate the cross-section as rectangular, neglecting the small sidewall angle. Nonetheless, the simulations match the experimentally deduced coupling strength. Neither electrostriction nor radiation pressure separately suffice to explain the experimental values of $\frac{2\gamma_{\text{SBS}}}{Q_{\text{m}}} \approx 12 \, \text{W}^{-1}\text{m}^{-1}$. These values are an order of magnitude larger than the state of the art, including on-chip chalcogenide\cite{Eggleton2013} and silicon nitride/silicon waveguides\cite{Shin2013b}.

In conclusion, we have demonstrated efficient interaction between near-infrared light and hypersound in a small-core silicon wire. The interaction is well described by the models, including phonon frequency, lifetime and coupling strength. There is ample scope for improving on these results. Currently limited by the pillar, the phonon lifetime may be increased by exciting asymmetric phononic modes \cite{Wilson-Rae2011} -- which are predicted to exhibit similar coupling strengths \cite{Qiu2013}. Then the pillar would barely vibrate, resulting in a large leakage reduction. The quality factor may also be improved by keeping the pillar but fully etching the wire in some regions. Alternatively, the intrinsic photon-phonon overlap could be increased by confining light to a narrow slot\cite{VanLaer2014}. Such ideas may enhance the Brillouin nonlinearity to a level sufficient for low-threshold lasing\cite{Li2014,Kabakova2013}, cascading\cite{Kang2009} or high-bandwidth fully non-resonant Brillouin scattering \cite{Pernice2009c}.

\vspace{4mm}

{\small
\textbf{Methods.} The cross-section (fig.1c) was milled by a focused ion beam. The platina around the silicon core is deposited for better visualization. We use the following abbreviations (fig.2): erbium-doped fiber amplifier (EDFA), band-pass filter (BPF), fiber polarization controller (FPC), intensity modulator (IM), electrical spectrum analyzer (ESA), light trap (LT), fiber Bragg grating (FBG) and photodetector (PD). The FBGs were a crucial part of our set-up. Produced by TeraXion Inc., these filters were custom-designed to have a flat response within the passband and drop to $-30 \, \text{dB}$ within $2.5 \, \text{GHz}$. We use the steep flank for filtering. Their bandwidth is $60 \, \text{GHz}$. In addition, we employ a pair of perfectly aligned FBGs for the gain experiment (fig.2b). On the theoretical side, we use the finite-element solver COMSOL to obtain the photonic and phononic modes. They were exported to MATLAB to calculate the coupling. Since our wires are aligned along a $\langle 110 \rangle$ axis, we rotated both the elasticity $(c_{11},c_{12},c_{44})=(166,64,79)\,\text{GPa}$ and the photoelasticity matrix $(p_{11},p_{12},p_{44})=(-0.09,0.017,-0.051)$ by $\pi/4$. In the simulations of the phonon leakage, we add an artificial silica matching layer with Young's modulus $\frac{i}{\zeta} E$ and density $-i\zeta\rho$. The layer absorbs incoming elastic waves without reflection. In a frequency-domain simulation, the quality factor can be found from $Q_{\text{m}}=\frac{\Re \Omega_{\text{m}}}{2 \Im \Omega_{\text{m}}}$. We optimize $\zeta$ for minimal $Q_{\text{m}}$. A typical value is $\zeta = 2$ for a $420 \, \text{nm}$-thick matching layer.

\textbf{Acknowledgement.} R.V.L. acknowledges the Agency for Innovation by Science and Technology in Flanders (IWT) for a PhD grant. This work was partially funded under the FP7-ERC-InSpectra programme and the ITN-network cQOM. R.V.L. thanks T. Van Vaerenbergh for reading the manuscript and L. Van Landschoot for taking SEM-pictures.

\textbf{Author contributions.} R.V.L. performed the fabrication, experiments, analysis and wrote the paper. B.K. gave experimental and conceptual advice. D.V.T. and R.B. supervised the work. All authors discussed the results and provided feedback on the manuscript.}

{\footnotesize
\bibliography{MyCollection_limited}
}
\section*{Supplementary information}

\subsection{Coupled-mode description of the gain experiments}

In this section we derive a simple model that captures the essential dynamics of forward Brillouin scattering in the presence of a background Kerr effect. Our analysis is very similar to earlier discussions of forward Brillouin scattering \cite{Kang2009,Rakich2012,Wang2011a} and Raman scattering \cite{Boyd2008}. Specifically, we describe under which circumstances forward Brillouin scattering can still be seen as a pure gain process. Thus the model includes
\begin{itemize}
\item cascading into higher-order Stokes and anti-Stokes waves,
\item four-wave mixing contributions from both the Brillouin and the Kerr effect and
\item the effect of these contributions on the SBS gain.
\end{itemize}
We assume that the electromagnetic field is composed of discrete lines, with $\tilde{A}_{n}(z)$ the complex amplitude of component $n$ with angular frequency $\omega_{n} = \omega_{0}+n\Omega$ at position $z$ along the guide. By definition, $\omega_{0}$ is the frequency of the pump. In the presence of weak nonlinear coupling between the waves, the evolution of the slowly-varying amplitudes is \cite{Boyd2008}
\begin{equation*}
\frac{d\tilde{A}_{n}}{dz} = -i \frac{\omega_{0}}{2cn_{\text{eff}}\epsilon_{0}} \tilde{P}^{\text{NL}}_{n}
\end{equation*}
with $\tilde{P}^{\text{NL}}_{n}$ the complex amplitude at frequency $\omega_{n}$ of the nonlinear polarization ${P}^{\text{NL}}(z,t) = \epsilon_{0} \chi^{\text{NL}}(z,t) A(z,t)$ with $A(z,t) = \tfrac{1}{2}\sum_{n} \tilde{A}_{n}(z) \exp{(i(\omega_{n}t - k_{n}z))} + \text{c.c.}$. Here we assumed that the cascading is limited to tens of higher-order sidebands, such that $\omega_{n} \approx \omega_{0}$ and that all components experience the same effective mode index $n_{\text{eff}}$. Further, the nonlinear susceptibility is given by
\begin{equation*}
\chi^{\text{NL}}(z,t) = 2n_{\text{eff}}\Delta n_{\text{eff}}(z,t)
\end{equation*}
in case the index changes $\Delta n_{\text{eff}}(z,t)$ are small. These index changes are composed of an instantaneous Kerr component and a delayed Brillouin component:
\begin{align*}
\Delta n_{\text{eff}}(z,t) &= \Delta n_{\text{eff,Kerr}}(z,t) +  \Delta n_{\text{eff,Brillouin}}(z,t) \\
			 	&=  \frac{\bar{n}_{2}}{A_{\text{eff}}}P(z,t) + \left.\frac{\partial n_{\text{eff}}}{\partial q}\right\vert_{q_{\text{avg}}} q(z,t)
\end{align*}
with $\bar{n}_{2}$ the nonlinear Kerr index averaged over the waveguide cross-section, $A_{\text{eff}}$ the effective mode area, $P(z,t)$ the total optical power and $q$ a coordinate describing the mechanical motion. There is a small shift in the average value of $q$ due to the constant component of the power $P(z,t)$. However, in practice this is very small such that effectively $q_{\text{avg}} \approx 0$. In addition, $\frac{\partial n_{\text{eff}}}{\partial q}$ is the sensitivity of the effective index with respect to motion. This factor contains contributions from both the moving boundary (radiation pressure) and the bulk (electrostriction). In this simple model, we characterize the mechanical mode as a harmonic oscillator in each cross-section $z$:
\begin{equation*}
\ddot{q}(z,t) + \Gamma_{\text{m}}\dot{q}(z,t) + \Omega^{2}_{\text{m}}q(z,t) = \frac{F(z,t)}{m_{\text{eff}}}
\end{equation*}
with $\frac{\Gamma_{\text{m}}}{2\pi}$ the Brillouin linewidth, $\Omega^{2}_{m} = \frac{k_{\text{eff}}}{m_{\text{eff}}}$ the angular frequency, $m_{\text{eff}}$ the effective mass of the mechanical mode per unit length and $F(z,t)$ the total force acting on that mode per unit length. Since this equation does not explicitly depend on $z$, $q(z,t)$ directly inherits its position-dependency from $F(z,t)$. Note that any propagation of phonons along the waveguide is neglected in this step. Each cross-section oscillates independently, reminiscent of the molecular vibration in Raman scattering \cite{Kang2009,Boyd2008}.

From power-conservation\cite{Rakich2009}, the optical force $F(z,t)$ per unit length can be related to $\frac{\partial n_{\text{eff}}}{\partial q}$ as
\begin{equation*}
F(z,t) = \frac{1}{c} \left.\frac{\partial n_{\text{eff}}}{\partial q}\right\vert_{q_{\text{avg}}} P(z,t)
\end{equation*}
The power $P(z,t) = 2 A^{2}(z,t)$ contains frequencies $n\Omega \, \,\forall n$ up to the total number of lines. However, we assume that only the component at $\Omega$ excites the mechanical motion. So we take $F(z,t) = \tfrac{1}{2}\tilde{F}_{\Omega} \exp{(i(\Omega t - K z))} + \text{c.c.}$ with $\tilde{F}_{\Omega} = \frac{1}{c}\frac{\partial n_{\text{eff}}}{\partial q}\tilde{P}_{\Omega}$ and $K = k_{0} - k_{-1} = \frac{\Omega}{v_{\text{g}}}$. The complex amplitude of the power $\tilde{P}_{\Omega}$ is given by
\begin{equation*}
\tilde{P}_{\Omega} = 2\sum_{n}\tilde{A}_{n}\tilde{A}^{*}_{n-1}
\end{equation*}
We normalized the amplitudes $\tilde{A}_{n}$ such that the power of wave $\omega_{n}$ is $|\tilde{A}_{n}|^{2}$. Thus the steady-state response of the harmonic oscillator is $\tilde{q} = Q_{\text{m}}\frac{\tilde{F}_{\Omega}}{k_{\text{eff}}}\mathcal{L}(\Omega)$ with the Lorentzian function $\mathcal{L}(\Omega) =\frac{1}{-2\Delta_r + i}$, the relative detuning $\Delta_r = \frac{\Omega - \Omega_{\text{m}}}{\Gamma_{\text{m}}}$ and the quality factor $Q_{\text{m}}=\frac{\Omega_{\text{m}}}{\Gamma_{\text{m}}}$. Therefore we can write the nonlinear index change in terms of the nonlinear Kerr and Brillouin parameters $\gamma_{\text{K}}$ and $\gamma_{\text{SBS}}$. Indeed, we have
\begin{align*}
\Delta \tilde{n}_{\text{eff}} &= \Delta \tilde{n}_{\text{eff,Kerr}} +  \Delta \tilde{n}_{\text{eff,Brillouin}} \notag \\
			 	&=  \frac{\gamma_{\text{K}}}{k_{0}}\tilde{P}_{\Omega} + \frac{\gamma_{\text{SBS}}}{k_{0}}\tilde{P}_{\Omega}\mathcal{L}(\Omega) \notag \\
				&= \frac{\tilde{P}_{\Omega}}{k_{0}}\gamma(\Omega) \notag
\end{align*}
where we defined the total nonlinearity parameter $\gamma(\Omega) = \gamma_{\text{K}}+\gamma_{\text{SBS}}\mathcal{L}(\Omega)$, using $\gamma_{\text{K}} \equiv k_{0}\frac{\bar{n}_{2}}{A_{\text{eff}}}$ and $\gamma_{\text{SBS}} \equiv \omega_{0}\frac{Q_{\text{m}}}{k_{\text{eff}}}\left(\frac{1}{c}\frac{\partial n_{\text{eff}}}{\partial q}\right)^{2}$. We note that this formula for the Brillouin nonlinearity is identical to the rigorous\cite{Rakich2012,Qiu2013,VanLaer2014} $\gamma_{\text{SBS}}=\omega_{\text{0}} Q_{\text{m}}|\langle \mathbf{f}, \mathbf{u}\rangle|^{2}/(4k_{\text{eff}})$ if we identify $\frac{1}{c}|\frac{\partial n_{\text{eff}}}{\partial q}| \equiv \frac{|\langle \mathbf{f}, \mathbf{u}\rangle|}{2}$. Hence the evolution of the amplitudes is
\begin{align}
& \frac{d\tilde{A}_{n}}{dz} = -i \frac{k_{0}}{2}\left( \Delta \tilde{n}_{\text{eff}} \tilde{A}_{n-1} +  \Delta \tilde{n}^{*}_{\text{eff}}\tilde{A}_{n+1} \right) \\
& \Delta \tilde{n}_{\text{eff}} = \frac{\tilde{P}_{\Omega}}{k_{0}}\gamma(\Omega) = \frac{2\gamma(\Omega)}{k_{0}}\sum_{n} \tilde{A}_{n}\tilde{A}^{*}_{n-1} \notag
\end{align}
These equations can be solved analytically since $\Delta \tilde{n}_{\text{eff}}$ turns out to be a constant of motion. Indeed, derivation yields
\begin{align*}
\frac{d\Delta \tilde{n}_{\text{eff}}}{dz} &\propto  \sum_{n}\left( \tilde{A}_{n}\frac{d\tilde{A}^{*}_{n-1}}{dz} + \frac{d\tilde{A}_{n}}{dz}\tilde{A}^{*}_{n-1}\right)  \\
 & \propto \sum_{n} \Delta \tilde{n}_{\text{eff}} \left( |\tilde{A}_{n}|^{2} - | \tilde{A}_{n-1}|^{2} \right) \\ 
& + \Delta \tilde{n}^{*}_{\text{eff}} \left(\tilde{A}_{n}\tilde{A}^{*}_{n-2} - \tilde{A}_{n+1}\tilde{A}^{*}_{n-1}\right) \\
&= 0
\end{align*}
Consequently, equation (1) can be solved either directly by using properties of the Bessel functions or indirectly by noting that $\Delta \tilde{n}_{\text{eff}}(z) = \Delta \tilde{n}_{\text{eff}}(0)$ such that the nonlinear interaction is equivalent to phase-modulation. Specifically,
\begin{align}
A(z,t) & = \frac{1}{2}\sum_{n} \tilde{A}_{n}(z)\exp{(i(\omega_{n}t - k_{n}z))} + \text{c.c.} \notag \\ 
& = \frac{1}{2} \exp{(-i k_{0} z \Delta n(z,t))} \times \notag \\
& \phantom{some text}  \sum_{n} \tilde{A}_{n}(0)\exp{(i(\omega_{n}t - k_{n}z))} + \text{c.c.}
\end{align}
Moreover, we have 
\begin{align}
\Delta n(z,t) &= |\Delta \tilde{n}_{\text{eff}}(0)|\sin{(\Omega t - Kz + \varphi_{0})} \\
		&= \frac{2|\gamma(\Omega)|}{k_{0}} \left|\sum_{n} \tilde{A}_{n}(0)\tilde{A}^{*}_{n-1}(0)\right| \sin{(\Omega t - Kz + \varphi_{0})} \notag
\end{align}
with $\varphi_{0} = \angle \left\{\Delta \tilde{n}_{\text{eff}}(0) \exp{(i \frac{\pi}{2})}\right\}$. As previously noted in the context of photonic crystal fibres\cite{Kang2009}, this is equivalent to phase-modulation with a depth $\xi$ determined by the strength of the input fields, the interaction length and the nonlinear parameter $|\gamma(\Omega)|$. The amplitudes of the individual components can finally be found by inserting $\exp{(i \xi \sin{\Phi})} = \sum_{n} \mathcal{J}_{n}(\xi) \exp{(i n \Phi)}$ with $\mathcal{J}_{n}$ the $n$th-order Bessel function of the first kind. To arrive at this phase-modulation picture, we assumed that all index changes originate from the beating at frequency $\Omega$. This is correct for the mechanical effect since it is weak off resonance. However, the Kerr response is non-resonant at telecom wavelengths. Thus its strength is the same at $\omega_{\text{0}} + n\Omega$ for all $n$. We include the $n\Omega$ ($n \neq 1$) Kerr-mediated coupling in the next paragraph, keeping in mind that equations (2)-(3) are only entirely correct when $\gamma_{\text{K}} = 0$.

To see how the modulation picture (2)-(3) relates to the traditional view of SBS as a pure gain process, we simplify equation (1) to the case of an undepleted pump, a Stokes and an anti-Stokes. Neglecting higher-order cascading, this yields
\begin{align}
& \frac{d\tilde{A}_{\text{s}}}{dz} = -i \gamma^{*}(\Omega) \left(|\tilde{A}_{\text{p}}|^{2} \tilde{A}_{\text{s}} + \tilde{A}^{2}_{\text{p}}\tilde{A}^{*}_{\text{as}}\right) \\ 
& \frac{d\tilde{A}_{\text{as}}}{dz}	=  -i \gamma(\Omega) \left(|\tilde{A}_{\text{p}}|^{2} \tilde{A}_{\text{as}} + \tilde{A}^{2}_{\text{p}}\tilde{A}^{*}_{\text{s}}\right)
\end{align}
In case $\tilde{A}_{\text{as}}(0)=0$, the initial evolution of the Stokes power is
\begin{equation*}
\frac{dP_{\text{s}}}{dz} = -2 \Im{\left\{\gamma(\Omega)\right\}}P_{\text{p}}P_{\text{s}}
\end{equation*}
Since $\Im{\left\{\gamma(\Omega)\right\}} = -\frac{\gamma_{\text{SBS}}}{4\Delta^{2}_{r}+1}$, we recover a Lorentzian Brillouin gain profile in this approximation:
\begin{align}
\frac{dP_{\text{s}}}{dz} &= G_{\text{SBS}}(\Omega)P_{\text{s}} \\
G_{\text{SBS}}(\Omega) &= \frac{2\gamma_{\text{SBS}}P_{\text{p}}}{4\Delta^{2}_{r}+1} \notag
\end{align}
Similarly, the anti-Stokes experiences a Lorentzian loss profile if $\tilde{A}_{\text{s}}(0)=0$. Thus the Kerr effect has no impact on the initial evolution of the Stokes wave. Therefore, forward SBS is a pure gain process as long as the anti-Stokes build-up is negligible. By numerically integrating equations (4) and (5), including linear losses, we confirm that this is the case in our experiments. The $n\Omega$ ($n \neq 1$) Kerr-mediated coupling does not change this conlusion. We can see this as follows. In the Lorentz-model for the permittivity, the Kerr response can be treated as a second-order nonlinear spring\cite{Boyd2008}

\begin{equation*}
\ddot{x} + \Gamma_{\text{e}}\dot{x} + \Omega^{2}_{e}(x)x = -\frac{e}{m_{\text{e}}}A
\end{equation*}
with $x$ the displacement of the electron cloud, $m_{\text{e}}$ the electron mass, $\Omega^{2}_{\text{e}}(x) = \frac{k_{\text{e}}(x)}{m_{\text{e}}}$ and $k_{\text{e}}(x) = k_{\text{e}}(0) + \frac{\partial^{2} k_{\text{e}}}{\partial x^{2}}x^{2}$ the nonlinear spring constant. Since $\omega_{n} \ll \Omega_{\text{e}}$, the oscillator responds instantaneously to the Lorentz-force $-eA$:

\begin{equation*}
\Omega^{2}_{e}(x)x = -\frac{e}{m_{\text{e}}}A
\end{equation*}
Thus the linear solution is $x_{\text{L}}(z,t) = \frac{-e}{k_{\text{e}}(0)}A(z,t)$. In the first Born approximation, the nonlinear displacement is

\begin{equation*}
x_{\text{NL}} = -\frac{1}{k_{\text{e}}(0)}\frac{\partial^{2} k_{\text{e}}}{\partial x^{2}}x_{\text{L}}^{3}
\end{equation*}
And the nonlinear polarization is ${P}^{\text{NL}} = \epsilon_{0} \chi^{\text{NL}} A = -N e x_{\text{NL}}$ with $N$ the atomic number density. This implies that the nonlinear polarization is proportional to $A^{3}(z,t)$. Unlike in the Brillouin case, the Lorentz oscillator does not filter out $0\Omega$, $2\Omega$, $3\Omega$, etc. terms. Selecting the right components of ${P}^{\text{NL}}$, we find that equations (4) and (5) are modified to
\begin{align*}
& \frac{d\tilde{A}_{\text{s}}}{dz} = -i \gamma^{*}(\Omega) \left(|\tilde{A}_{\text{p}}|^{2} \tilde{A}_{\text{s}} + \tilde{A}^{2}_{\text{p}}\tilde{A}^{*}_{\text{as}}\right) -i \gamma_{\text{K}} |\tilde{A}_{\text{p}}|^{2} \tilde{A}_{\text{s}}\\ 
& \frac{d\tilde{A}_{\text{as}}}{dz}	=  -i \gamma(\Omega) \left(|\tilde{A}_{\text{p}}|^{2} \tilde{A}_{\text{as}} + \tilde{A}^{2}_{\text{p}}\tilde{A}^{*}_{\text{s}}\right)  -i \gamma_{\text{K}} |\tilde{A}_{\text{p}}|^{2} \tilde{A}_{\text{as}}
\end{align*}
for a strong, undepleted pump. The added terms on the right generate a constant phase shift and do, therefore, not alter the conclusion that these equations yield Brillouin gain when $\tilde{A}_{\text{as}}(0)=0$. However, such added terms do invalidate the phase-modulation solution (2)-(3).

Back to that solution (2)-(3), at first sight we expect a Fano-like resonance for the Stokes power because the modulation depth depends on $|\gamma(\Omega)|$ and not on $\Im{\left\{\gamma(\Omega)\right\}}$. However, the input phase $\varphi_{0}$ also contains phase information on $\gamma(\Omega)$. We analytically check that the phase-modulation picture is equivalent to a pure gain process in the low-cascading regime. Combining equations (2) and (3) with only an initial pump and Stokes wave, we find
\begin{equation*}
\tilde{A}_{\text{s}}(z) = \tilde{A}_{\text{s}}(0) - \mathcal{J}_{1}\left(\xi\right)\tilde{A}_{\text{p}}(0)\exp{(-i(\varphi_{0} + \pi))}
\end{equation*}
with $\xi = 2|\gamma(\Omega)|\sqrt{P_{\text{s}}(0)P_{\text{p}}}z$ the unitless cascading parameter. The power of the Stokes wave then becomes
\begin{equation*}
P_{\text{s}}(z) = P_{\text{s}}(0)\left(1 - 2 \Im{\left\{\gamma(\Omega)\right\}}P_{\text{p}}z\right) + \frac{\xi^{2}}{4}P_{\text{p}}
\end{equation*}
Here we approximated the Bessel function as $\mathcal{J}_{1}\left(\xi\right) \approx \frac{\xi}{2}$, which is valid in the low-$\xi$ regime. The last term, containing $\xi^{2}$, gives rise to a Fano-resonance but is smaller than the other terms in this regime. Taking the derivative and letting $z \rightarrow 0$, we indeed recover the gain equation (6). In our experiments we reach values of $\xi \approx 0.4$ in the longest waveguides and at maximum pump power. To conclude, we can safely neglect higher-order cascading and treat forward SBS as a pure gain process driven exclusively by the Brillouin nonlinearity. In the presence of linear optical losses, the modified evolution of the Stokes wave is
\begin{align*}
\frac{dP_{\text{s}}}{dz} = & \left(G_{\text{SBS}}(\Omega)\exp{\left(-\alpha z\right)} -\alpha \right)P_{\text{s}} \notag \\
& G_{\text{SBS}}(\Omega) = \frac{2\gamma_{\text{SBS}}P_{\text{p}}}{4\Delta^{2}_{r}+1} \notag
\end{align*}
with $\alpha$ the linear optical loss and $P_{\text{p}}$ the input pump power. The analytical solution of this equation is
\begin{align*}
P_{s}(L) = & P_{s}(0) \exp{\left(G_{\text{SBS}}(\Omega) L_{\text{eff}} - \alpha L \right)}
\end{align*}
with $L_{\text{eff}} = \frac{1 - \exp{\left(- \alpha L \right)}}{\alpha}$ the effective interaction length. In the case of nonlinear losses $\alpha(P_{\text{p}})$ the equations can be integrated numerically.

\subsection{Coupled-mode description of the cross-phase modulation experiments}
\begin{figure*}
\centering
\label{fig:2}
\subfigure{
\begin{tikzpicture}[>=stealth']
		
   		 \node[anchor=south west,inner sep=0] (image) at (0,0) {\raisebox{5pt}{\includegraphics[width=0.6\columnwidth]{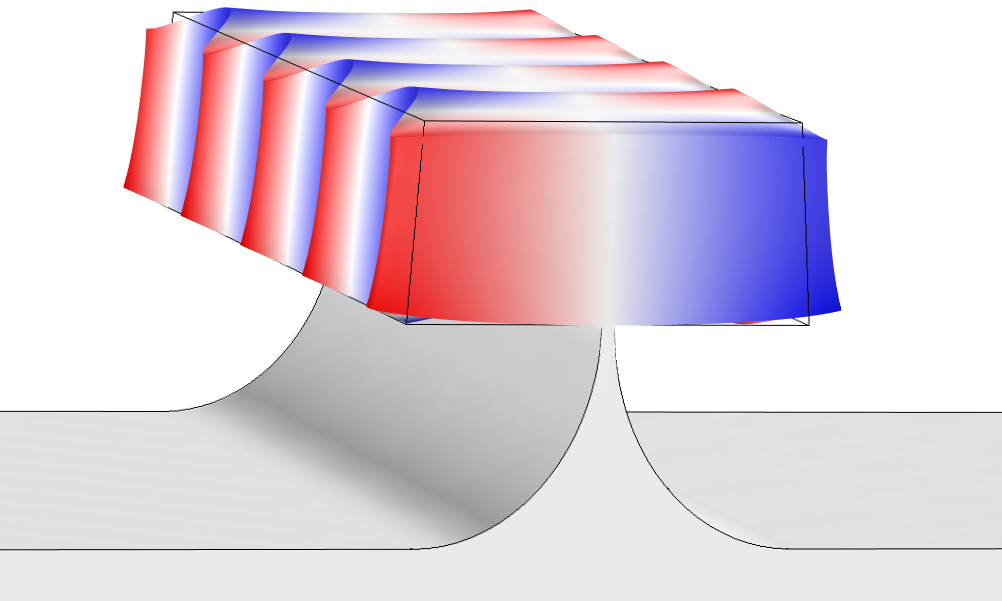}}};

		\node at (-0.7,-0.4) {\large{\textbf{\textcolor{white}{a}}}};
		\node at (-0.3,-0.08) {\large{\textbf{\textcolor{black}{a}}}};
    		
\end{tikzpicture}
}
\subfigure{
\hspace{2mm}

\beginpgfgraphicnamed{fig:SBSgain}
\begin{tikzpicture}[>=stealth']
\node at (0.25,-3.2) {\large{\textbf{b}}};
\draw[black,thick] (0,0) rectangle (1.5,0.8) ;
\draw[red,thick] (0.1,0.4) -- (1.2,0.4);
\filldraw[red](1.2,0.4) circle (0.7mm);
\foreach \n in {0,...,11} {\draw[red,rotate around={360/11*(\n-1):(1.2,0.4)}] (1.2,0.4) -- (1.45,0.4);};
\foreach \n in {0,...,11} {\draw[red,rotate around={(360/11*(\n-1)+360/22):(1.2,0.4)}] (1.2,0.4) -- (1.38,0.4);};
\node at (0.34,0.56) {{$\lambda_{P}$}};
\node at (0.75,-0.2) {1550 \hspace{-0.5mm}{nm}};

\draw[blue,very thick] (1.5,0.4) -- (2,0.4);
\draw[thick] (2,0.3) rectangle (2.2,0.5);
\draw[thick] (2,0.5) -- (2.2,0.3);
\node at (2.45,0.18) {\scriptsize{50\%}};

\draw[blue,very thick] (2.1,0.3) -- (2.1,-1.3);
\draw[blue,very thick] (2.08,-1.29) -- (2.6,-1.29);
\begin{scope}[rotate around={90:(3.0,-1.06)},shift={(0.6,1.12)}]
\draw[thick] (2.7,-1.16) circle (0.1);
\draw[thick] (3.0,-1.06) circle (0.2);
\draw[thick] (3.3,-1.16) circle (0.1);
\end{scope}
\begin{scope}[shift={(2.28,-0.48)}]
\node[rotate=90] at (0.0,0) {\scriptsize{FPC}};
\end{scope}

\begin{scope}[shift={(-1.3,0.0)}]
\node at (4.3,-1.3) {\scriptsize{IM}};
\draw[thick] (3.9,-1.6) rectangle (4.7,-1.0);
\draw[very thick,brown] (4.3,-0.7) -- (4.3,-0.99);
\draw[thick] (4.3,-0.5) circle (0.2);
\node at (4.3,-0.71) {\huge$\tilde{}$};
\node at (4.80,-0.21) {\scriptsize{10 \hspace{-0.5mm}{GHz}}};

\draw[blue,very thick] (4.7,-1.3) -- (5.2,-1.3);
\draw[thick] (5.5,-1.3) circle (0.3);
\draw[->,thick] (5.5,-1.3) ++(140:2.1mm) arc (-220:40:2.1mm) --++(120:1mm);
\draw[thick] (5.45,-0.73) -- (5.45,-0.67) -- (5.55,-0.67) -- (5.55,-0.73);
\node at (5.72,-0.64) {\tiny{LT}};
\draw[blue,very thick] (5.5,-0.99) -- (5.5,-0.7);
\draw[blue,very thick] (5.81,-1.3) -- (6.31,-1.3);

\draw[thick] (6.32,-1.5) -- (6.32,-1.1);
\draw[thick] (6.38,-1.5) -- (6.38,-1.1);
\draw[thick] (6.44,-1.5) -- (6.44,-1.1);
\draw[thick] (6.50,-1.5) -- (6.50,-1.1);
\draw[blue,very thick] (6.51,-1.3) -- (7.0,-1.3);
\node at (6.41,-0.95) {\scriptsize{FBG}};

\draw[thick] (7.0,-1.5) -- (7.0,-1.1) -- (7.4,-1.3) -- (7.0,-1.5) -- cycle;
\node at (7.2,-0.95) {\scriptsize{EDFA}};
\draw[blue, very thick] (7.38,-1.3) -- (7.91,-1.3);
\draw[thick] (7.91,-1.5) rectangle (8.41,-1.1);
\node at (8.16,-0.95) {\scriptsize{BPF}};
\draw[thick] (8.0,-1.45) .. controls (8.15,-1.05) .. (8.3,-1.45);
\end{scope}

\draw[blue, very thick] (7.11,-1.3) -- (7.63,-1.3) --  (7.63,-2.8);
\draw[thick] (7.51,0.3) rectangle (7.71,0.5);
\draw[thick] (7.51,0.3) -- (7.71,0.5);

\begin{scope}[rotate around={-90:(3.0,-1.06)},shift={(1.0,4.83)}]
\draw[thick] (2.7,-1.16) circle (0.1);
\draw[thick] (3.0,-1.06) circle (0.2);
\draw[thick] (3.3,-1.16) circle (0.1);
\end{scope}
\begin{scope}[shift={(7.45,-2.08)}]
\node[rotate=-90] at (0.0,0) {\scriptsize{FPC}};
\end{scope}
\node[] at (7.3,-3.5) {\scriptsize{PD}};
\node at (7.85,-3.25) {\scriptsize{1\%}};
\node at (8.56,-2.75) {\scriptsize{99\%}};
\draw[blue, very thick] (7.73,-2.9) -- (9.4,-2.9) -- (9.4,-2.4);

\draw[thick] (7.53,-3.0) rectangle (7.73,-2.8);
\draw[thick] (7.53,-2.8) -- (7.73,-3.0);
\begin{scope}[rotate around = {180:(7.61,0.6)},shift={(-0.02,3.7)}]
\draw[blue,very thick] (7.61,0.5) -- (7.61,1.0);
\draw[thick,fill,black!90] (7.51,1.0) .. controls (7.56,1.2) and (7.66,1.2) .. (7.71,1.0);
\end{scope}

\draw[blue!40,fill,opacity=0.3,rounded corners] (3.84,-2.3) rectangle (5.4,-0.53);
\node[] at (4.65,-1.9) {\scriptsize{Remove $\lambda_{P}$}};
\node[] at (4.87,-2.15) {\scriptsize{and $\lambda_{AS}$}};

\draw[red!30,fill,opacity=0.2,rounded corners] (1.65,-2.35) rectangle (5.45,-0.03);
\node[] at (2.55,-2.1) {\scriptsize{Create Stokes}};

\draw[blue,very thick] (2.2,0.4) -- (3.1,0.4);
\draw[thick] (3.1,0.2) -- (3.1,0.6) -- (3.5,0.4) -- (3.1,0.2) -- cycle;
\draw[blue,very thick] (3.48,0.4) -- (4.4,0.4);
\node at (3.3,0.73) {\scriptsize{EDFA}};

\draw[thick] (4.4,0.2) rectangle (4.9,0.6);
\node at (4.65,0.73) {\scriptsize{BPF}};
\draw[thick] (4.5,0.25) .. controls (4.65,0.65) .. (4.8,0.25);

\draw[blue,very thick] (4.91,0.4) -- (7.51,0.4);
\draw[thick] (5.9,0.52) circle (0.1);
\draw[thick] (6.2,0.62) circle (0.2);
\draw[thick] (6.5,0.52) circle (0.1);
\node[] at (6.22,0.22) {\scriptsize{FPC}};
\node at (7.65,0.13) {\scriptsize{50\%}};

\draw[blue,very thick] (7.61,0.5) -- (7.61,0.75);
\draw[thick,fill,black!90] (7.51,0.75) .. controls (7.56,0.95) and (7.66,0.95) .. (7.71,0.75);
\node[] at (7.95,0.85) {\scriptsize{PD}};
\draw[blue,very thick] (7.71,0.4) -- (8.3,0.4);

\draw[thick] (8.6,0.4) circle (0.3);
\draw[->,thick] (8.6,0.15) ++(140:2.1mm) arc (220:-40:2.1mm) --++(-120:1mm);
\draw[blue,very thick] (8.6,0.1) -- (8.6,-0.6) -- (8.2,-0.6);

\begin{scope}[shift={(1.7,0.7)}]
\draw[thick] (6.32,-1.5) -- (6.32,-1.1);
\draw[thick] (6.38,-1.5) -- (6.38,-1.1);
\draw[thick] (6.44,-1.5) -- (6.44,-1.1);
\draw[thick] (6.50,-1.5) -- (6.50,-1.1);
\node at (6.41,-0.95) {\scriptsize{FBG}};
\end{scope}
\draw[blue,very thick] (8.01,-0.6) -- (7.6,-0.6);
\begin{scope}[rotate around={90:(7.55,0.7)},shift={(-1.36,-0.1)}]
\draw[thick,fill,black!90] (7.51,0.75) .. controls (7.56,0.95) and (7.66,0.95) .. (7.71,0.75);
\end{scope}
\node[] at (7.54,-0.38) {\scriptsize{PD}};
\draw[blue!40,fill,opacity=0.2,rounded corners] (7.3,-1.1) -- (7.3,-0.1) -- (8.25,-0.1) -- (8.25,0.75) -- (8.95,0.75) -- (8.95,-1.1) -- (7.3,-1.1) -- cycle;
\node[] at (8.15,-0.96) {\scriptsize{Detect $\lambda_{S}$}};

\draw[blue,very thick] (8.9,0.4) -- (9.4,0.4);

 \node[anchor=south west,inner sep=0,rotate=0] (image) at (9.2,-0.95) {\raisebox{0pt}{\includegraphics[scale=0.06]{img/waveguide_view4.png}}};

\draw[black!30,fill,opacity=0.2,rounded corners] (9.1,-1.08) rectangle (10.4,0.14);
\draw[white,<->,very thin,decorate,decoration={snake,amplitude=.2mm,segment length=1mm,post length=1mm,pre length=1mm}] (9.53,-0.2) -- (9.64,-0.65);
\begin{scope}[shift={(10.24,-0.5)}]
\node[rotate=-90] at (0.0,0) {\scriptsize{Chip}};
\end{scope}

\draw[rotate around={2.5:(9.483,0.06)}] (9.418,0.10) -- (9.544,0.10) -- (9.481,-0.005) -- (9.418,0.10) -- cycle;
\draw[very thin] (9.44,0.05) arc[start angle=150,end angle=40,x radius=0.5mm,y radius = 0.5mm];
\draw[very thin] (9.46,0.035) arc[start angle=150,end angle=40,x radius=0.25mm,y radius = 0.25mm];
\draw[thin] (9.4,0.4) .. controls (9.65,0.4) and (9.45,0.25) .. (9.48,0.12);

\begin{scope}[rotate around={180:(9.483,0.06)},shift={(-0.19,1.05)}]
\draw[rotate around={2.5:(9.483,0.06)}] (9.418,0.10) -- (9.544,0.10) -- (9.481,-0.005) -- (9.418,0.10) -- cycle;
\draw[very thin] (9.44,0.05) arc[start angle=150,end angle=40,x radius=0.5mm,y radius = 0.5mm];
\draw[very thin] (9.46,0.035) arc[start angle=150,end angle=40,x radius=0.25mm,y radius = 0.25mm];
\end{scope}

\draw[blue,very thick] (9.4,-1.3) -- (9.4,-1.8);
\draw[thin] (9.68,-1.05) .. controls (9.695,-1.25) and (9.38,-1.15) .. (9.4,-1.3);
\draw[<->,thin] (9.6,0.3) .. controls (9.65,0.15) .. (9.6,-0.1);
\begin{scope}[shift={(-0.07,-1)}]
\draw[<->,thin] (9.6,0.3) .. controls (9.55,0.15) .. (9.6,-0.1);
\end{scope}

\begin{scope}[rotate around={90:(5.8,-1.3)},shift={(-0.5,-3.6)}]
\draw[thick] (5.5,-1.3) circle (0.3);
\draw[->,thick] (5.5,-1.3) ++(140:2.1mm) arc (-220:40:2.1mm) --++(120:1mm);
\draw[blue,very thick] (5.5,-0.99) -- (5.5,-0.5);
\end{scope}
\begin{scope}[rotate around={90:(7.55,0.7)},shift={(-2.86,-1.1)}]
\draw[thick,fill,black!90] (7.51,0.75) .. controls (7.56,0.95) and (7.66,0.95) .. (7.71,0.75);
\end{scope}
\node[] at (8.59,-1.89) {\scriptsize{PD}};

\end{tikzpicture}
\endpgfgraphicnamed
}
\caption{\textbf{Characterization of the backward Brillouin scattering.} \textbf{a}, Propagating version of the Fabry-P\'{e}rot phononic mode (see fig.1d for comparison). \textbf{b}, Experimental set-up used to observe the backward SBS gain. This time the Stokes and pump wave counterpropagate through the chip, exciting phonons that satisfy $K = 2k_{0}$.}
\end{figure*}
In the cross-phase modulation (XPM) experiments, we study the phase modulation imprinted on a probe wave by a strong intensity-modulated pump. The pump and its sidebands are located at frequencies $\omega_{0}$, $\omega_{1} = \omega_{0} + \Omega$ and $\omega_{-1} = \omega_{0} - \Omega$. The probe has frequency $\omega_{\text{pr}}$. The four-wave mixing interaction between these waves imprints sidebands $\omega_{\text{pr}}\pm \Omega$ on the probe. We monitor the power $P_{\text{imprint}}$ in the $\omega_{\text{imprint}} = \omega_{\text{pr}} + \Omega$ sideband at the end of the waveguide as a function of $\Omega$.

If there were only Brillouin coupling between the waves, the effective index would be modulated exclusively at frequency $\Omega$. However, the Kerr effect responds equally well to the beat notes $\Delta_{0} = \omega_{0} - \omega_{\text{pr}}$ and $\Delta_{-1} = \omega_{-1} - \omega_{\text{pr}}$. So there are four pathways to $\omega_{\text{imprint}}$: 
\begin{align*}
\omega_{\text{imprint}} &= \omega_{\text{pr}} + \left( \omega_{1} - \omega_{0} \right) \\
\omega_{\text{imprint}} &= \omega_{\text{pr}} + \left( \omega_{0} - \omega_{-1} \right) \\
\omega_{\text{imprint}} &= \omega_{1} - \Delta_{0} \\
\omega_{\text{imprint}} &= \omega_{0} - \Delta_{-1}
\end{align*}
Both the Kerr and the Brillouin effect take the first two, but only the Kerr effect takes the latter two pathways. Therefore the Kerr effect manifests itself with double strength in these experiments. Building on the formalism of supplementary section A, we calculate the imprinted sideband power $P_{\text{imprint}}$. The index modulation is
\begin{align}
\Delta n(z,t) &= |\Delta \tilde{n}_{\text{eff},\Omega}|\sin{(\Omega t - Kz + \varphi_{\Omega})} \\
		&+ |\Delta \tilde{n}_{\text{eff},\Delta_{0}}|\sin{(\Delta_{0} t - \left(k_{0} - k_{\text{pr}}\right)z + \varphi_{\Delta_{0}})} \notag \\
		&+ |\Delta \tilde{n}_{\text{eff},\Delta_{-1}}|\sin{(\Delta_{-1} t - \left(k_{-1} - k_{\text{pr}}\right)z + \varphi_{\Delta_{-1}})} \notag
\end{align}
with the following definitions
\begin{align*}
\Delta \tilde{n}_{\text{eff},\Omega} &= \frac{\tilde{P}_{\Omega}}{k_{0}}\left\{\gamma_{\text{K}}+\gamma_{\text{SBS}}\mathcal{L}(\Omega)\right\}\\
\Delta \tilde{n}_{\text{eff},\Delta_{0}} &= \frac{\tilde{P}_{\Delta_{0}}}{k_{0}}\gamma_{\text{K}} \\
\Delta \tilde{n}_{\text{eff},\Delta_{-1}} &= \frac{\tilde{P}_{\Delta_{-1}}}{k_{0}}\gamma_{\text{K}}
\end{align*}
As before, we denote the angles $\varphi = \angle \left\{\Delta \tilde{n}_{\text{eff}} \exp{(i \frac{\pi}{2})}\right\}$. We also define a modulation depth $\xi = k_{0}z|\Delta \tilde{n}_{\text{eff}}|$ for each beat note. Next, we insert equation (7) in equation (2) and apply the Bessel expansion $\exp{(i \xi \sin{\Phi})} = \sum_{n} \mathcal{J}_{n}(\xi) \exp{(i n \Phi)}$ to each of the beat notes. This results in
\begin{align}
A(z,t)  &= \frac{1}{2}\sum_{klm} \mathcal{J}_{k}(\xi_{\Omega})  \mathcal{J}_{l}(\xi_{\Delta_{0}})  \mathcal{J}_{m}(\xi_{\Delta_{-1}}) \times \notag \\ 
& \phantom{}   \exp{(-i(k \Phi_{\Omega} + l \Phi_{\Delta_{0}} + m \Phi_{\Delta_{-1}})} \times \notag \\
&  \sum_{n} \tilde{A}_{n}(0)\exp{(i(\omega_{n}t - k_{n}z))} + \text{c.c.}
\end{align}
Only three terms in the Bessel expansion influence $P_{\text{imprint}}$ when $\xi$ is small. In particular, for $(klm)=(-100)$, $(010)$ and $(001)$ the frequencies $\omega_{\text{pr}}$, $\omega_{1}$ and $\omega_{0}$ are shifted to $\omega_{\text{imprint}}$ respectively. Working out equation (8) for these terms, we obtain
\begin{align*}
\tilde{A}_{\text{imprint}}(z) &= -\frac{\xi_{\Omega}}{2}\exp{(i \varphi_{\Omega})}\tilde{A}_{\text{pr}} + \frac{\xi_{\Delta_{0}}}{2}\exp{(-i \varphi_{\Delta_{0}})}\tilde{A}_{\text{1}} \\
& \quad + \frac{\xi_{\Delta_{-1}}}{2}\exp{(-i \varphi_{\Delta_{-1}})}\tilde{A}_{\text{0}}
\end{align*}
for the amplitude $\tilde{A}_{\text{imprint}}(z)$ of the imprinted tone. Here we used $\mathcal{J}_{1}\left(\xi\right) = \frac{\xi}{2}$ for small $\xi$. Since the beat note amplitudes are $\tilde{P}_{\Omega} = 2\left(\tilde{A}_{1}\tilde{A}^{*}_{0} + \tilde{A}_{0}\tilde{A}^{*}_{-1}\right)$, $\tilde{P}_{\Delta_{0}} = 2 \tilde{A}_{0}\tilde{A}^{*}_{\text{pr}}$ and $\tilde{P}_{\Delta_{-1}} = 2 \tilde{A}_{-1}\tilde{A}^{*}_{\text{pr}}$, we finally obtain
\begin{align*}
P_{\text{imprint}}(z) &=  |\gamma_{\text{XPM}}(\Omega)|^{2} |\tilde{P}_{\Omega}|^{2} P_{\text{pr}}\frac{z^{2}}{4}
\end{align*}
with $\gamma_{\text{XPM}}(\Omega) = 2\gamma_{\text{K}}+\gamma_{\text{SBS}}\mathcal{L}(\Omega)$. Therefore we use the Fano lineshape $|\frac{\gamma_{\text{XPM}}(\Omega)}{2\gamma_{\text{K}}}|^{2}$ as a fitting function for the normalized probe sideband power.

\subsection{Measurement of backward scattering}
\begin{figure*}
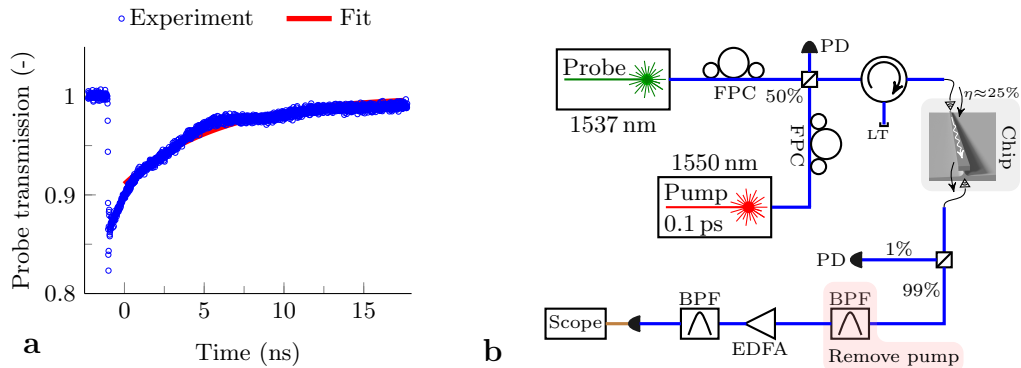

\centering
\label{fig:2}
\subfigure{
%
%
%
%

\endpgfgraphicnamed
}
\\
\caption{\textbf{Measurement of the free-carrier lifetime.} \textbf{a}, Oscilloscope trace of the probe power. The pump pulse arrives at $t=-1 \, \text{ns}$. We start the fit a nanosecond later to avoid fitting to photodiode ringing artifacts. \textbf{b}, Pump-probe set-up used to obtain the trace. The band-pass filter (BPF) has more than $50 \, \text{dB}$ extinction at $1550 \, \text{nm}$.}
\end{figure*}
So far we focused on forward SBS, in which the excited phonons have a very short wavevector $K = \frac{\Omega}{v_{\text{g}}}$ because the pump and Stokes have nearly equal wavevectors. However, the phononic mode demonstrated in this work (fig.1d) can also be operated at another point in its dispersion diagram. When the Stokes and pump counterpropagate through the wire, they generate the class of phonons that obey $K = 2k_{0}$. The propagating version (fig.5a) of the Fabry-P\'{e}rot mechanical mode (fig.1d) may then induce gain as well. We reconfigured our gain experiment (fig.5b) to study such modes.

We also find a Lorentzian gain profile (fig.2a), but this time at $13.7 \, \text{GHz}$. A fit yields $Q_{\text{m}}=971$. This propagating phononic mode exhibits a strongly reduced photon-phonon coupling. From the experiment, we find that $\frac{2\gamma_{\text{SBS}}}{Q_{\text{m}}} = 0.37 \, \text{W}^{-1}\text{m}^{-1}$: a factor $30$ lower than in the forward case. We attribute this reduction to destructive interference between electrostriction and radiation pressure, as predicted before\cite{Qiu2013} for fully suspended wires. Because of this low overlap, we observe the backward resonance only in the long spirals.

From our finite-element models, we expect this propagating mode (fig.5a) at $14.4 \, \text{GHz}$ with a coupling of $\frac{2\gamma_{\text{SBS}}}{Q_{\text{m}}} = 0.41 \, \text{W}^{-1}\text{m}^{-1}$. Therefore we suspect that this is indeed the observed mode. Further investigations should resolve this issue, as the simulations predict that there are propagating modes with better coupling at higher frequencies. These frequencies were not accessible in our current set-up.

%

\subsection{Measurement of the free-carrier lifetime}

Free electrons and holes, created by two-photon absorption (TPA) in our experiments, induce significant free-carrier absorption (FCA) and free-carrier index changes (FCI) above a certain power threshold. As reflected in the saturation of the SBS gain (fig.3a), this threshold is about $25 \, \text{mW}$ in our $450 \times 230 \, \text{nm}$ silicon wires. From the observations
\begin{itemize}
\item that our finite-element and coupled-mode modeling of the Brillouin effect matches the experiments
\item and that the off-resonance background is flat in the XPM experiment
\end{itemize}
we have evidence that the free-carriers are not noticeably influencing our results below the threshold. Nevertheless, we performed a cross-FCA experiment (fig.6b) to exclude the possibility of a significant drop in free-carrier lifetime $\tau_{\text{c}}$ caused by the underetch of our wires.

The pump was a $\approx 100 \, \text{fs}$-pulse with a repitition rate of $\frac{1}{50 \, \text{ns}}$ and peak power of $\approx 1 \, \text{kW}$. When a pump pulse arrives, it creates many free-carriers by TPA. The free-carriers recombine before the next pump pulse arrives. Their presence is read out by monitoring the power of a c.w. probe wave on a high-speed oscilloscope. Thus the transmission $T$ of the probe is
\begin{equation*}
T = \exp{\left(-\alpha_{\text{FCA}}(t)\right)} = \exp{\left(-\alpha_{\text{FCA}}(t_{0})\exp{\left(-\frac{t}{\tau_{\text{c}}}\right)}\right)}
\end{equation*}
where we normalized the transmission to the case without FCA. Here we exploited the relation $\alpha_{\text{FCA}}(t) \propto N(t)$ with $N(t)$ the free-carrier concentration.

The experiments yield typical values of $\tau_{\text{c}} = 6.2 \, \text{ns}$ before the etch and $\tau_{\text{c}} = 5.7 \, \text{ns}$ after the etch in identical waveguides. Hence there is, if at all, only a minor decrease of $\tau_{\text{c}}$ due to the underetch. The associated bandwidth of $f_{3\,\text{dB}}=\tfrac{1}{2\pi\tau_{\text{c}}}= 28 \, \text{MHz}$ suggests a negligible FCI-effect at $10 \, {\text{GHz}}$. As a precaution, we work when possible -- in the longer wires -- with low power (below $15 \, \text{mW}$ on-chip) in the XPM-experiments. The free-carrier nonlinearity $\gamma_{\text{FCI}}$ can, in principle, always be reduced below $\gamma_{\text{K}}$ because $\gamma_{\text{FCI}} \propto P_{\text{pump}}$ while $\gamma_{\text{K}}$ does not depend on $P_{\text{pump}}$.

%
%


\end{document}